# Magnetoelectric-field electrodynamics: Search for magnetoelectric point scatterers

E. O. Kamenetskii

Department of Electrical and Computer Engineering,
Ben Gurion University of the Negev, Beer Sheva, Israel

September 11, 2022

**Abstract**
Resonant scattering of electromagnetic (EM) waves by small particles is considered as one of the basic problem in metamaterial science. At present, special subwavelength resonators are considered as structural elements in chiral and bianisotropic metamaterials. There is a general consensus that these small scatterers behave like "artificial atoms" – meta-atoms – with strong electrical and magnetic responses and an interconnection between these responses. However, the observed effect of magnetoelectric (ME) coupling in these meta-atoms is not associated with the near-field manipulation properties caused by intrinsic magnetoelectricity. This arises the question whether ME point scatterers of EM radiation really exist. In this paper, we show that there are mesoscopic structures with electric and magnetic dipole-carrying excitations that behave like point scatterers with their inherent magnetoelectricity. In such subwavelength resonators, coherent oscillations of the electric polarization and magnetization can be considered as quasistatic oscillations described by electrostatic (ES) and magnetostatic (MS) scalar wave functions. The ME resonance effect arises from the coupling of two, ES and MS, oscillations. The near fields of these resonators, called the ME near fields, are characterized by simultaneous violation of time reversal and inversion symmetry. In study of ME fields and EM problems associated with these fields, we put forward the concept of ME-field electrodynamics. Like the axion-field electrodynamics and the axion-polariton problem, we are talking about modified Maxwell equations with a presudoscalar source terms and the quantum field theory for massive particles. While the axion fields are boson scalar fields, the ME fields are fermionic fields. These fields are characterized by energy eigenstates with rotational superflows and quantized vortices.

## I. INTRODUCTION

The relationship between electromagnetism and magnetoelectricity is a subject of a strong interest in microwave and optical wave physics and material sciences. Electromagnetism is based on the fact that a moving charge has a magnetic field. The magnetic field in electromagnetic processes is controlled by electric current, macroscopic or microscopic. Magnetoelectricity is associated with the structures in which the polarization and magnetization are coupled in some way. One of the interesting questions about whether electromagnetism and magnetoelectricity can coexist without an extension of Maxwell's theory arises when we study EM wave scattering from *subwavelength resonant objects* with assumed ME properties. This is related to the fundamental problems of ME interactions in metamaterials and polaritons. It can also refer to ME phenomena associated with the concepts of axion electrodynamics.

Confining the EM radiation on a strongly subwavelength scale is one of the key elements that make it possible to achieve a high interaction strength of the field-matter interaction. An intuitive physical paradigm is that the propagation of a wave is not significantly affected by subwavelength objects. However, a widespread aspiration in wave physics is to be able to affect at scales that are much smaller than the wavelengths. Furthermore, free-space coupling to strongly subwavelength individual optical elements is a central theme in quantum optics, as it allows to control and manipulate the properties of quantum systems. For objects that are very small compared to the incident wavelength, the EM *non-retarded approximation* becomes applicable. Usually, such subwavelength scatterers are treated as dipoles or a set of dipoles. The value of subwavelength resonators is that they can be used as building blocks for large, complex structures. One example that can exhibit a variety of exotic and useful properties of these subwavelength resonators is metamaterials [1]. Currently, subwavelength resonators are considered as structural elements in chiral and bianisotropic metamaterials. It is assumed that these individual scatterers behave as *"artificial atoms"* or, in other words, meta-atoms with strong *electric and magnetic responses*, and there is a general consensus that these scatterers should be described by ME point-dipole interactions [2 – 8]. In the analysis of ME interactions in bianisotropic metamaterials, simple models with dipolar terms are used. The known inverse-scattering-problem solutions [9, 10] and retrieval via measurements of the scattering-matrix characteristics [11 – 12], used for the parameter reconstruction of bianisotropic samples, are *far-field techniques*. However, it is not possible to find 3D *subwavelength details* of a ME scatterer retrieved from far-field data. Known models of small electric and magnetic dipole antennas do not shed light on the nature of the ME polarizability, since the near-field *energy of ME coupling* (cross-polarization energy) is not considered. The main point is that there are no solutions of Maxwell's equations with two local sources, electric and magnetic currents, which are supposedly linked by electromagnetic forces. The realization of local coupling between electric and magnetic dipoles is associated with the violation of both spatial and temporal inversion symmetries. In the subwavelength (quasistatic) vacuum region, this is impossible from the point of view of classical electrodynamics. The "first-principle", "microscopic-scale" ME effect of a structure composed by "glued" pairs of electric and magnetic dipoles raise questions on the ways of local probing the dynamic ME parameters, since the 3D near-field structure of such a probe should violate both the spatial and temporal inversion symmetries. It is known that far-field retrieved permittivity and permeability frequently retain non-physical values in the regions of the material resonances where most interesting features are expected. Far-field retrieved parameters of dipolar cross-polarization ME resonances retain much greater non-physical value. When the basic concepts of bianisotropy are analyzed from the EM far-field characteristics, we just only have the far-field "*illusion of bianisotropy*".

The question about ME polarizability and the near-field manipulation arises also in the studies of EM effects of chirality. Electromagnetically, chiral samples are often considered as a particular case of bianisotropic samples. To intuitively explain the optical responses of small chiral particles, the dipole approximation is used. A point-like chiral system is described by the dynamic polarizability tensor, which is applied to study chiral optomechanical and scattering processes. Since chiroptical effects are usually hampered by weak chiral light-matter interaction, it is argued that to enhance the chiral effects it is necessary that the near field remains chiral in the process. This is an attempt to enhance the near-field intensity while preserving the ME properties. Based on an analysis of the interaction of chiral light and chiral specimens, new mechanisms of enantiomer discrimination and separation in optics have been proposed. Different plasmonic, dielectric, and semiconductor nanostructures have recently been proposed as a viable route for near-field enhancement of chiral light-matter interactions [13 – 23]. Review articles on these problems may be helpful to the reader [24, 25]. Investigations of the EM effects of chirality are being advanced by new ideas and concepts. In particular, there is an idea to use the strong interaction of light and matter in nanophotonic resonators based on the concept of quasinormal modes [26]. In a view of discussions on the problem on locality in the chiral light-matter interaction, new mechanisms have been proposed for controlling the chiral light–matter interaction. In this analysis, the degree of chirality of locally chiral EM fields is defined by the minimal overlap between the electric-field vector and its own mirror twin [27 – 29]. However, in all of the above works, the effects of *propagating waves* of the EM fields are used to characterize the properties of chiral particles. Widely studied superchiral fields for enantioselective molecular sensing [13], magnetoelectric fields that can exist locally due to the interference of several monochromatic circularly polarized plane waves [30], the optical electromagnetic helicity conservation theorem [31] - all these physical phenomena using to the analysis of chiral objects can be observed only on the basis of far-field experiments.

A question on the *evanescent-wave* characteristics of bianisotropic and enantiomeric "artificial atoms" remains open. Without knowing the near field structure of such a 3D-confined subwavelength samples, a deep understanding of electrodynamics of bianisotropic and enantiomeric metamaterials is still missing. The observed phenomena can hardly be associated with the near-field manipulation effects caused by *intrinsic magnetoelectricity*. Constitutive elements of these metamaterials are made of matter without any microscopic or mesoscopic properties of a natural ME effect. It is evident that 3D-confined subwavelength samples with intrinsic ME resonances should be characterized by the ME energy. It is known that as volumes smaller than the wavelength are probed, measurements of EM energy of usual (non-ME) samples become uncertain, highlighting the difficulty with performing measurements in this regime. There is Heisenberg's uncertainty principle binding the electric and magnetic fields of the EM wave [32]. This fact gives us much greater uncertainty in local probing of the fields of the ME point scatterer. In a subwavelength region, the coupling characteristics of such near fields can be represented as the structures of cross $\vec{E} \times \vec{B}$ and dot $\vec{E} \cdot \vec{B}$ products of the electric and magnetic fields. The effect of magnetoelectricity manifests itself in the simultaneous time reversal and inversion symmetry breakings. The ME near fields of a subwavelength sample should be characterized by a certain *pseudoscar parameters*. Moreover, supposing that in a subwavelength region both structures of cross $\vec{E} \times \vec{B}$ and dot $\vec{E} \cdot \vec{B}$ products exist, one should assume the presence of helicity properties of the fields. There should be rotating fields with spin and orbital angular momenta.

The intrinsic ME polarizability is observed in multiferroics – the materials in which both ferroelectric and magnetic orders are present. Fundamental microscopic excitations that exhibit the electric and magnetic dipole moments in multiferroics, are electromagnons. ME coupling of the electromagnon resonance are the magnon excitations accompanied by the magnetic and ferroelectric orders. In controlling of light with electromagnons, the effects of optical activity and asymmetric propagation arise due to dynamic ME susceptibility that cross-couples the electric and magnetic fields of light [33 – 37]. However, to the best of our knowledge, no "artificial atoms" – the 3D-confined subwavelength resonators – made of such a multiferroic matter, are used as structural element of metamaterials. What could be the ways for realizing subwavelength resonant elements with intrinsic ME polarizability and ME near fields? To answer this question, we should focus our attention on two problems. The first problem is about *3D-confined polaritons*. The second problem concerns study of ME fields in the view of the *axion electrodynamics* aspects.

The ability to focus EM radiation far beyond the diffraction limit has enabled numerous applications. A primary strategy to achieve such subwavelength light confinement is to use polaritons. Polaritons are structures with strong interaction of photon with *dipole-carrying excitations* in materials. The energy stored in the wave is partitioned among the driving fields and the internal degrees of freedom excited by the incident electric and magnetic fields. Various types of polaritonic light–matter effects are observed. They differ in resonant responses of the materials. Among the most discussed types of polaritons there are phonon polaritons, exciton polaritons, surface-plasmon polaritons, and magnon polaritons [38, 39]. Recently, there are studies of polaritons generated in magnetoelectric multiferroics. Since fundamental excitations of this type of material consist of spin waves and lattice vibrations, one may generate magnon polaritons and phonon polaritons simultaneously [40, 41]. The dipole-carrying excitations propagate with wavelengths that are much smaller than those of free-space photons of the same energy. Polaritons are characterized by a wavevector that lies beyond the light line. The lower branches of polaritons are dark by this criterion and do not couple to free space photons because of the *momentum mismatch problem*. In the case of plasmon polaritons, the field-matter coupling is through the electric dipole interaction. Electric fields are confined to the interface on a scale that is determined by a small wavelength of plasmon oscillations. The electric field is enhanced. In the case of magnon polaritons, the field-matter coupling is through the magnetic dipole interaction. Magnetic fields are confined to the interface on a scale that is determined by a small wavelength of magnon oscillations. The magnetic field is enhanced.

In the above cases of plasmon and magnon polaritons, 2D structures are used and no subwavelength resonators are considered. Such 3D-confined effects are possible only for the case of ME polaritons, where the field-matter coupling is through both the electric and magnetic dipoles and *both the electric and magnetic fields are enhanced*. This is the *photon trapping effect*. The only way to focus EM radiation far beyond the diffraction limit in a 3D-confined subwavelength region is to use power-flow vortices defined by cross $\vec{E} \times \vec{B}$ field products in a near-field region. In this case, we are not talking on the momentum mismatch problem, but about the *angular momentum mismatch problem*. By the angular momentum conservation, the spin and orbital momenta of the dipole-carrying excitations in a sample should be coupled to the spin and orbital angular momenta of EM radiation in the subwavelength near-field region. It is worth noting that in the case of plasmon polaritons or magnon polaritons, the momentum matching is in the field region near the interface. This is a local (quasistatic) matching. At the same time, the *matching of the spin and orbital angular momenta* in the case of ME polaritons is *global*. It means that the angular momentum matching is implemented globally, involving both the near- and far-field regions. This integration will lead to an extremely strong light-matter interaction. We term such phenomena the ME-field electrodynamics. In ME-field electrodynamics, we observe photons with *curved wave fronts of circularly polarized waves*. These are *twisted polaritons*. It can be postulated that the effect of strong interaction of light with a 3D-confined subwavelength ME sample occurs when near fields are twisted. Another aspect, which should also be noted, concerns the fact that the realization of the light-matter interactions in this case requires that internal dynamical processes in the sample are characterized by breaking both the spatial and time-reversal symmetry.

We call a subwavelength resonator with an intrinsic ME effect an *intrinsic-ME resonator (IMER)*. A special attention should be paid to ME cavity polariton, when a subwavelength IMER is placed inside a reflective cavity. In this case, strong coupling regime is due to the formation of entangled eigenstates of dipolar excitations in an IMER and EM radiation confined in a reflective cavity. When a reflective cavity is nearly resonant with a ME oscillation, trapped photons may be emitted and reabsorbed multiple times before being lost to dissipation or cavity leakage. Absorption and re-emission of photons in the cavity give rise to light–matter mixed eigenstates. Photons become strongly interacting and condensation of photons can be observed. To certain extent, such effects look similar to cavity exciton polaritons [42 – 52]. There are, however, fundamental differences, which we will further discuss in this paper.

Coherent oscillations of electric polarization and magnetization in a 3D-confined subwavelength IMER are viewed as quasistatic oscillations with electrostatic (ES) and magnetostatic (MS) scalar wave functions. *Coupling of two, ES and MS, oscillations is considered as the effect of magnetoelectricity*. Near fields of an IMER – the ME near fields – are generated by simultaneous time reversal and inversion symmetry breakings. For EM radiation, subwavelength regions with quasistatic magnetoelectric ME resonances, are regions of singular states. Our purpose is to study ME fields and EM problems associated with these fields. Particular attention should be paid to the fact that the ME coupling observed in a subwavelength resonator is a relativistic effect of rotating fields. Like the axion-field electrodynamics and the axion-polariton problem [53 – 62], we are talking about modified Maxwell equations with a presudoscalar source term and the quantum field theory for massive particles. Axions arise out of pseudoscalar fields derived from the Chern–Simons theory in condensed matter physics.

One of the effective ways to enhance the interaction of EM radiation with matter is to bound both dipole-carrying excitations and photons to a 3D-confined subwavelength resonant region. In this paper, it is shown that at a proper geometry and boundary conditions of a subwavelength resonator, one observes coherent oscillations of electric polarization and magnetization. We show that mesoscopic structures with electric and magnetic dipole-carrying excitations can behave as point scatterers with intrinsic magnetoelectricity. The effect of magnetoelectricity is due to coupling of two, ES and MS, concurrent orders generated by simultaneous time reversal and inversion symmetry breakings. For EM radiation, subwavelength regions with quasistatic magnetoelectric ME resonances, are regions of singular states. These singularities are found in a variety of forms. The near-fields of open quasistatic ME resonators are topological-nature fields – the ME fields. Our purpose is to study ME fields and EM problems associated with these fields. While the axion fields are boson scalar fields, the ME fields are fermionic fields. These fields are characterized by energy eigenstates with rotational superflows and quantized vortices. There the field structures of cross $\vec{E} \times \vec{B}$ and dot $\vec{E} \cdot \vec{B}$ products of the electric and magnetic fields in a subwavelength vacuum region.

## II. QUASISTATIC RESONANCES IN SUBWAVELENGTH SAMPLES

In our search for the IMER point scatterer, the main concepts are as follows. We do not apply the model of a scatterer with coupled electric and magnetic dipoles, which is widely used in numerous studies of metamaterials. We also do not consider the effects of ME polarizability of hypothetical "artificial atoms" made of a multiferroic material. We assume that in our case, the specially shaped specimen is made of a lossless solid material that has the properties of electric polarization and magnetization, but is not possessed of microscopic intrinsic magnetoelectricity. We are addressing to an analysis of a mesoscopic effect of *dynamical magnetoelectricity* in a subwavelength sample, which originates from *coupled MS and ES resonances*. Herewith MS resonances arise due to magnetization dynamics and ES resonances appear due to electric-polarization dynamics in the material. It is supposed that in the resonance states of coupled MS and ES oscillations, a violation of both spatial and temporal inversion symmetry occurs. Any EM retardation effects are disregarded. It means that no Mie resonances are taken into account. This is the case of scattering of a small particle at the quasistatic limit ($ka \approx 0$, where $k$ is an EM-wave wavenumber in a medium and $a$ is a characteristic size of a particle).

The MS and ES resonances can be observed in a temporally dispersive regime, when the light frequency is close to the resonance frequency of dipole-carrying excitations in the material. In these resonances, a continuum approach is applied and quantum confinement effects are considered based on the concept that the MS function $\psi$ ($\vec{H} = -\vec{\nabla}\psi$) and the ES function $\phi$ ($\vec{E} = -\vec{\nabla}\phi$) are *scalar wave functions*. Each of the scalar wave functions is characterized by a module and a phase. In the spectral problem solutions, MS and ES wave functions play a role of order parameters for interacting dipoles. The coupling states of two, MS and ES, concurrent orders are considered as the ME states. The strongest effect of the field-matter interaction is observed when these internal ME resonances are characterized by energy quantization with discrete energy levels. While the quasistatic oscillations are not identified with the scales of individual atoms, they can be represented as dynamics of quasiparticles of the elementary dipole-dipole excitations. Scalar wave functions are the generating functions of the vector fields. For electromagnetic radiation, quasistatic ME resonances caused by dipolar excitations in a subwavelength sample are dark states. In this section, we present an initial formal analysis of MS and ES oscillations in a subwavelength resonator, considering these oscillations as topological quantum states.

**A. MS resonances in subwavelength samples**

In a subwavelength 3D sample of a magnetic insulator with strong temporal dispersion of permeability $\ddot{\mu}(\omega)$, magnetic dipole oscillations can be considered, neglecting the time variations of the energy of the electric field as compared to the time variations of the energy of the magnetic field. His is a case of MS resonances. A system of differential equations for the electric and magnetic fields is without the electric displacement current:

$$\nabla \cdot \vec{B} = 0 \tag{1}$$

$$\vec{\nabla} \times \vec{H} = 0 \tag{2}$$

$$\vec{\nabla} \times \vec{E} = -\frac{\partial \vec{B}}{\partial t} \tag{3}$$

When the electric displacement current is excluded, $\frac{\partial \vec{D}}{\partial t} = 0$, we have $\vec{H} = -\vec{\nabla}\psi$. A wave equation for the MS-wave function is given as

$$\vec{\nabla} \cdot \left[ \ddot{\mu} \left( \vec{\nabla}\psi \right) \right] = 0. \tag{4}$$

From the Fraday law (3), we write

$$\vec{\nabla} \times \frac{\partial \vec{E}}{\partial t} = \frac{\partial^2 \vec{B}}{\partial t^2}. \tag{5}$$

Assuming the sample is a dielectric with an isotropic permittivity, we have $\frac{\partial^2 \vec{B}}{\partial t^2} = 0$. So, it follows that the magnetic field in small resonant objects vary linearly with time. This leads to arbitrary large fields at early and late times and is excluded on physical grounds. An evident conclusion suggests itself at once: the magnetic field in MS resonances is a constant quantity. This contradicts the resonant conditions of dynamic processes. Thus, it becomes clear that in a proposed subwavelength resonator with magnetic-dipole oscillations the curl electric field arising from the Faraday law is unphysical [63]. This lack of field symmetry arises as a significant issue if we consider the problem of the interaction of MS resonances with an external electromagnetic field when analyzing subwavelength scatterers.

Nevertheless, "electromagnetic democracy" can be someway restored if we assume that the magnetic-dipole dynamics in a magnetic insulator sample is accompanied with the induced electric polarization characterized by temporal-dispersion permittivity tensor $\ddot{\varepsilon}_{ind}(\omega)$. In such an assumption, $\vec{D}_{ind} = \ddot{\varepsilon}_{ind}(\omega)\vec{E}$ and, obviously, $\frac{\partial^2 \vec{B}}{\partial t^2} \neq 0$ in Eq. (5). So, the fact of neglect of the electric displacement current does not contradict the presence of the curl electric field. A suitable way to induce the electric polarization in a subwavelength magnetic sample can be a topological effect associated with *chiral magnetic currents* at the sample boundary. We can say that the basic effect for observing MS resonances in subwavelength 3D samples of a magnetic insulator is because of an interaction between the ferromagnetic order subsystem and the electric polarization subsystem, which arise due to chiral magnetic currents on the surface of the sample.

In analyses of MS-wave problems in a magnetic insulator, Eq. (4) is known as Walker's equation [64]. It is worth noting that in the literature, where the problems on the long-range magnetization oscillations and waves in ferrite samples are studied, one can see that in the spectral problems formulated exceptionally for the MS-potential wave function $\psi$, no consideration is given to the electric fields arising from Faraday's law [65 – 69]. However, in our study of MS resonances in subwavelength samples, the role of an electric fields is fundamental.

**B. ES resonances in subwavelength samples**

In a subwavelength 3D sample of a dielectric material with strong temporal dispersion of permittivity $\ddot{\varepsilon}(\omega)$, electric dipole oscillations can be considered. With these oscillations, we neglect the time variations of the energy of the magnetic field as compared to the time variations of the energy of the electric field. In such a case, the electromagnetic duality is broken. We have a system of differential equations for the electric and magnetic fields without the magnetic displacement current:

$$\nabla \cdot \vec{D} = 0 \qquad (6)$$

$$\vec{\nabla} \times \vec{E} = 0 \qquad (7)$$

$$\vec{\nabla} \times \vec{H} = \frac{\partial \vec{D}}{\partial t} \qquad (8)$$

When the magnetic displacement current is excluded, $\frac{\partial \vec{B}}{\partial t} = 0$, we have $\vec{E} = -\vec{\nabla}\phi$. A wave equation for the ES-wave function is written as

$$\vec{\nabla} \cdot \left[ \vec{\vec{\varepsilon}} \left( \vec{\nabla}\phi \right) \right] = 0. \tag{9}$$

Formally, it can be assumed that for a certain geometric shape of a subwavelength dielectric sample, solutions of the boundary volume problem based on Eq. (9) are the spectral solutions of ES oscillations.

From the Ampere-Maxwell law (8), we write

$$\vec{\nabla} \times \frac{\partial \vec{H}}{\partial t} = \frac{\partial^2 \vec{D}}{\partial t^2}. \tag{10}$$

Suppose that the sample has no magnetic anisotropy. In this case, assuming that the magnetic displacement current is zero, we have $\frac{\partial^2 \vec{D}}{\partial t^2} = 0$. So, it follows that the electric field in small resonant objects vary linearly with time. This leads to arbitrary large fields at early and late times and is excluded on physical grounds. An evident conclusion suggests itself at once: the electric field in ES resonances is a constant quantity. This contradicts the resonant conditions of dynamic processes. Thus, it becomes clear that in a proposed subwavelength resonator with electric-dipole oscillations the curl magnetic field arising from the Ampere-Maxwell law is unphysical [70].

Nevertheless, the problem of the "electromagnetic democracy" in ES resonances can be solved if we assume that electric-dipole dynamics in a dielectric sample is accompanied with the *induced magnetization* characterized by temporal-dispersion permeability $\vec{\vec{\mu}}_{ind}(\omega)$. In such an assumption, $\vec{B}_{ind} = \vec{\vec{\mu}}_{ind}(\omega)\vec{H}$ and, obviously, $\frac{\partial^2 \vec{D}}{\partial t^2} \neq 0$ in Eq. (10). So, the fact of neglect of the magnetic displacement current does not contradict the presence of the curl magnetic field if the magnetization is induced.

A suitable way to induce the magnetization in a dielectric can be a topological effect associated with *chiral electric currents* at the sample boundary. In this regard, it is appropriate to refer to structures of topological insulators (TI). In particular, it was shown that the edge of a 2D TI supports counterpropagating chiral states with opposite spin projections on the axis perpendicular to the plane of the sample. These edge modes serve as conducting channels that give rise to the quantum spin Hall effect. As a result, ferromagnetic ordering can be observed [71, 72]. Another effect of induced magnetization is shown in a composite structure in which the side surface of a 3D TI is covered by a magnetic specimen with ferromagnetic order [73]. However, these effects of magnetization in TI have no connection to our study of ES resonances. In this paper, we show that ES resonances can be observed in non-topological magnetic dielectric. In a subwavelength sample made of a magnetic insulator, an interaction between the ferromagnetic order subsystem and the electric polarization subsystem can occurs. In this case, ES resonances are detected in the presence of MS resonances.

## III. CONFINEMENT EFFECTS IN MS AND ES OSCILLATIONS

Neglecting any EM retardation processes in a subwavelength sample, we assume, at the same time, that the dimensions of the sample are large enough to ignore the microscopic effects associated with the dipole-dipole interactions. When the characteristic size of a sample is of the same magnitude as the wavelength of the scalar wave functions, the confinement effects can be observed. In this case, we are talking about resonances of scalar wave functions: the MS-wave function $\psi$ and the ES-wave function $\phi$. To describe these resonant effects of electric or magnetic dipolar oscillations, the Schrodinger-like equations for functions $\psi$ and $\phi$ should be used. Assuming that the MS and ES wave functions satisfy, respectively, Eqs. (4) and (9) inside a sample, outside the sample there should be Laplace equations for functions $\psi$ and $\phi$. To solve wave equations (4) and (9), the Neumann-Dirichlet (ND) boundary conditions for scalar wave functions and their derivates should be applied at the boundaries of the samples. The physically observable results are obtained when the eigenstates of the boundary volume problem for functions $\psi$ and $\phi$ form a complete-set orthonormal basis. These effects of quantum confinement must be related to discrete energy levels.

However, it turns out that in some specific geometry of a magnetic sample and a specific form of the permeability tensor $\ddot{\mu}(\omega)$, the EM boundary condition used for the vector $\vec{B}$ $\left(\text{where } \vec{B} = -\ddot{\mu}(\vec{\nabla}\psi)\right)$ differ from the ND boundary conditions when solving the boundary volume problem for function $\psi$. Similarly, in some specific geometry of a dielectric sample and a specific form of the permittivity tensor $\ddot{\varepsilon}(\omega)$, the EM boundary condition used for the vector $\vec{D}$ $\left(\text{where } \vec{D} = -\ddot{\varepsilon}(\vec{\nabla}\phi)\right)$ does not correspond to the ND boundary conditions necessary for solving the boundary volume problem for function $\phi$. It can be shown that the discussed above chiral currents at the boundary of sample are related to the differences between the ND and EM boundary conditions when $\ddot{\mu}(\omega)$ and $\ddot{\varepsilon}(\omega)$ are *gyrotropic* tensors. While analyzing what is a suitable geometrical shape of a 3D-confined subwavelength sample where topological edge currents can occur, we should not consider the spherical sample, since the sphere is simply connected and, thus, every current loop can be contracted on the surface to a point. Obviously, there should not be also an ellipsoid. At the same time, a quasi-2D disk turns out to be the most suitable sample shape. When the current loop is on the lateral surface of such a disk, we have a non-simply connected domain with topologically protective edge currents. This non-simply connected region obviously has a local field-flux. With the separation of variables for scalar wave functions in a quasi-2D geometry, one can analytically solve the boundary volume problem of quasistatic oscillations.

Both MS and ES resonances are observed with quantum confinement effects for scalar wave function $\psi$ and $\phi$. In a case of MS resonances, the fact that MS potential function $\psi$ is introduced does not contradict the presence of the curl electric field if we assume that magnetic-dipole dynamics in a magnetic sample is accompanied with the induced polarization. Also, for ES resonance, the fact that the ES potential function $\phi$ is introduced does not contradict the presence of the curl magnetic field if we assume that electric-dipole dynamics in a dielectric sample is accompanied with the induced magnetization.

In MS resonances, constitutive equation for magnetic materials $\vec{B} = \mu_0\left(\vec{H} + \vec{m}\right)$ gives from Eq. (1):

$$\vec{\nabla} \cdot \vec{H} = -\vec{\nabla} \cdot \vec{m} \qquad (11)$$

We know that in macroscopic electrodynamics of a magnetic insulator, there are three types of current densities: the density of the electric displacement current $\varepsilon_0 \frac{\partial \vec{E}}{\partial t}$, the electric current density arising from polarization $\frac{\partial \vec{p}}{\partial t}$, and the divergenceless electric current density arising from magnetization $\vec{\nabla} \times \vec{m}$ [74]. Following the Helmholtz decomposition of any vector field into potential and curl parts [75], we have $\vec{m} = \vec{m}_{pot} + \vec{m}_{curl}$, where $\vec{\nabla} \times \vec{m}_{pot} = 0$ and $\vec{\nabla} \cdot \vec{m}_{curl} = 0$. The potential part of the magnetization defines the solutions for the magnetic scalar potential ($\vec{H} = -\vec{\nabla} \psi$). The curl part of magnetization plays the role of an effective electric current density. Since $\vec{\nabla} \cdot \vec{B} = 0$ holds, we can define $\vec{B}$ in terms of a magnetic vector potential. From the "full" Maxwell equation (with the electric displacement current), we can write

$$\vec{\nabla} \times \vec{B} = \mu_0 \vec{\nabla} \times (\vec{H} + \vec{m}) = \mu_0 \left( \varepsilon_0 \frac{\partial \vec{E}}{\partial t} + \frac{\partial \vec{p}}{\partial t} + \vec{\nabla} \times \vec{m} \right). \qquad (12)$$

When the electric displacement current is zero, $\frac{\partial (\varepsilon_0 \vec{E} + \vec{p})}{\partial t} = 0$, we identically have $\vec{\nabla} \times \vec{H} = 0$, that is $\vec{H} = -\vec{\nabla} \psi$. If, from the other hand, we assume that

$$\left| \frac{\partial (\varepsilon_0 \vec{E} + \vec{p})}{\partial t} \right| \ll |\vec{\nabla} \times \vec{m}|, \qquad (13)$$

we also have: $\vec{\nabla} \times \vec{H} \approx 0$ and thus $\vec{H} \approx -\vec{\nabla} \psi$. An inequality (13) implies that variations of electric energy are negligibly small. Assuming that the permittivity of a magnetic insulator is sufficiently high, we can rewrite Eq. (13) as

$$\left| \frac{\partial \vec{p}}{\partial t} \right| \ll |\vec{\nabla} \times \vec{m}|. \qquad (14)$$

Eq. (14) means that in the dynamical process, the lines of the vector field of the electric-dipole moments $\vec{p}$ are *"frozen"* in the lines of the magnetization field $\vec{\nabla} \times \vec{m}$. Such a situation can be observed when the vector $\vec{p}$ *rotates synchronously* with the vector $\vec{m}$ in the laboratory coordinate system. In a quasi-2D disk of a magnetic insulator, vector $\vec{m}$ has both spin and orbital angular momenta. In this case, vector $\vec{p}$ must also have spin and orbital angular momenta. So, together with the precession of $\vec{m}$ we will observe the effect of the vector $\vec{p}$ precession. The electric field vector will also rotate synchronously with vector $\vec{m}$ [76 – 79].

For ES resonances, with use of the constitutive equation for dielectric material $\vec{D} = \varepsilon_0 \vec{E} + \vec{p}$, we have from Eq. (6)

$$\varepsilon_0 \vec{\nabla} \cdot \vec{E} = -\vec{\nabla} \cdot \vec{p}. \tag{15}$$

We also can use the Helmholtz decomposition $\vec{p} = \vec{p}_{pot} + \vec{p}_{curl}$, where and $\vec{\nabla} \times \vec{p}_{pot} = 0$ and $\vec{\nabla} \cdot \vec{p}_{curl} = 0$. The potential part of the polarization in a dielectric defines the solutions for the electric scalar potential ($\vec{E} = -\vec{\nabla}\phi$). At the same time, the curl part of polarization field is not related to the curl electric field [74]. It turns out, however, that in the case of ES resonances in a subwavelength dielectric sample, the situation can be quite different. Since $\vec{\nabla} \cdot \vec{D} = 0$ holds, we can formally define $\vec{D}$ in terms of an electric vector potential $\vec{D} \equiv \vec{\nabla} \times \vec{C}$. Taking into account the constitutive equation (15) and the "full" Maxwell equation (with the magnetic displacement current), we have:

$$\vec{\nabla} \times \vec{D} = \vec{\nabla} \times \left(\varepsilon_0 \vec{E} + \vec{p}\right) = -\varepsilon_0 \mu_0 \frac{\partial \vec{H}}{\partial t} + \vec{\nabla} \times \vec{p}. \tag{16}$$

When the magnetic displacement current is zero, $\frac{\partial \vec{H}}{\partial t} = 0$, we identically have $\vec{\nabla} \times \vec{E} = 0$, that is $\vec{E} = -\vec{\nabla}\phi$. From the other hand, we can suppose that for the terms of the Eq. (16) the following inequality holds:

$$\varepsilon_0 \mu_0 \left|\frac{\partial \vec{H}}{\partial t}\right| \ll \left|\vec{\nabla} \times \vec{p}\right|. \tag{17}$$

As a result, we also obtain: $\vec{\nabla} \times \vec{E} \approx 0$, that is $\vec{E} \approx -\vec{\nabla}\phi$. But, along with this, inequality (17) may mean that in a dynamical process in the subwavelength dielectric sample, vector $\vec{H}$ *rotates quite synchronously* with vector $\vec{\nabla} \times \vec{p}$ in the lab coordinate system. So, in the rotational coordinate system, the lines of the magnetic field $\vec{H}$ are "frozen" in the lines of $\vec{\nabla} \times \vec{p}$. Obviously, such a situation can take place when both the magnetic field and electric polarization vectors have the spin and orbital angular momenta. Since, following Eq. (15), $\vec{\nabla} \cdot \vec{p} \neq 0$, the quantity $\vec{\nabla} \times \vec{p}$ cannot be observed locally in a sample. But it can be observed globally via circulation on contour $\mathcal{L}$ on the lateral surface of a quasi-2D disk. In other words, $\vec{\nabla} \times \vec{p}$ is physical only after integrating around a closed path. The integral of the curvature of $\vec{\nabla} \times \vec{p}$ defined over the contour $\mathcal{L}$ is a topological invariant which can be physically measured as the quanta of the orbital polarization. This model predicts the topological magnetoelectric effect, where an edge current of orbital polarization generates a topological contribution to the magnetization. The current loop on the lateral surface of a quasi-2D disk is a non-simply connected domain with topologically protective electric edge currents. This non-simply connected region is obviously related to the magnetic-field flux. As a result, we have $\vec{B} = \ddot{\mu}_{ind}(\omega)\vec{H}$ in a dielectric sample.

## IV. SPECTRAL CHARACTERISTICS OF FERRITE MAGNETOELECTRIC SCATTERERS

Inequality (13) shows the obvious fact that in a sample with MS resonances, variation of density of electric energy is negligibly small. Similarly, inequality (17) shows that in a sample with ES resonances, variation of density of magnetic energy is negligibly small. Nevertheless, the processes of transition of magnetic energy into electric energy and vice versa, that is, the property of dynamic magnetoelectricity, can be observed when the MS and ES resonances become globally coupled due to orbital chiral currents.

The effects of orbital magnetization and orbital electric polarization manifest themselves in a subwavelength quasi-2D disk, made of a magnetic insulator – yttrium iron garnet (YIG). In such a sample, magnetization is the primary order parameter, which can induce electric polarization dynamics. So, the ES resonances are detected in the presence of the MS resonances. On the other hand, the spectral characteristics of the MS resonances are determined by the ES resonances. In a quasi-2D ferrite disk, one can use separation of variables and the "particle in a box" model for the modes of the MS scalar wave function $\psi(\vec{r},t)$. These modes, called magnetic dipolar modes (MDM), are due to magnetic dipole-dipole interaction in the ferromagnetic order subsystem. Oscillations caused by electric dipole-dipole interaction in the electric polarization subsystem we call electric dipolar modes (EDM). These oscillations, described by the ES scalar wave function $\phi(\vec{r},t)$, are considered as an extension of MDM oscillations.

### A. MDM oscillations in the magnetic subsystem

The theory of spectral properties of MDMs in a ferrite-disk resonator is published in [79 – 85]. Here we provide brief excerpts from this theory, emphasizing the main aspects needed for the current study on spectral properties of ferrite magnetoelectric scatterers.

When analyzing MDM oscillations in a ferrite disk, two types of solutions for scalar wave function $\psi(\vec{r},t)$ are considered. There are the spectral solutions of the energy eigenstates, conventionally called the *G*-modes solutions, and the spectral solutions of the power-flow-confinement states, conventionally called the *L*-modes solutions [77 – 85]. For *G* modes, we define the *energy eigenstates* of MS oscillations based on the Schrödinger-like equation for scalar wave function $\psi(\vec{r},t)$ with use of the ND boundary conditions. In case of *L* modes, we consider normalization to the *power-flow density*

$$\vec{\mathcal{J}} = \frac{i\omega}{4}\left(\psi\vec{B}^* - \psi^*\vec{B}\right) \tag{18}$$

using the EM boundary conditions. Obviously, when characterizing the MDM oscillations, the resonant states of the *G* and *L* modes should be considered together.

Analyzing the *G*-mode solutions in a cylindrical coordinate system $(z, r, \theta)$, we determine the membrane function $\tilde{\eta}(r,\theta)$ by the Bessel-function order and the number of zeros of the Bessel function corresponding to the radial variations. Membrane functions $\tilde{\eta}(r,\theta)$ is a single-valued function. On a lateral surface of a ferrite disk, the ND boundary conditions for mode *n* are written as

$$\left(\tilde{\eta}_n\right)_{r=\mathcal{R}^-} - \left(\tilde{\eta}_n\right)_{r=\mathcal{R}^+} = 0 \tag{19}$$

and

$$\mu\left(\frac{\partial \tilde{\eta}_n}{\partial r}\right)_{r=\mathcal{R}^-} - \left(\frac{\partial \tilde{\eta}_n}{\partial r}\right)_{r=\mathcal{R}^+} = 0, \tag{20}$$

where $\mathcal{R}$ is a disk radius. For *G* modes, the spectral problem gives the energy orthogonality relation: $(E_n - E_{n'})\int_{S_c} \tilde{\eta}_n \tilde{\eta}_{n'}^* dS = 0$. The quantity $E_n$ is considered as density of accumulated magnetic energy of mode *n*. This is the average (on the RF period) energy accumulated in a flat ferrite-disk region of in-plane cross-section and unit length along *z* axis. Since the space of square integrable functions is a Hilbert space with a well-defined scalar product, we can introduce a basis set. The mode amplitude can be interpreted as the probability to find a system in a certain state *n*. Using the principle of wave-particle duality, one can describe this oscillating system as a collective motion of quasiparticles. There are "flat-mode" quasiparticles at a reflexively-translational motion behavior between the lower and upper planes of a quasi-2D disk. Such quasiparticles are called "light" magnons. In our study we consider "light" magnons in ferromagnet as quanta of collective MS spin waves that involves the precession of many spins on the long-range dipole-dipole interactions. It is different from the short-range magnons for exchange-interaction spin waves with a quadratic character of dispersion. The meaning of the term "light", used for the condensed MDM magnons, arises from the fact that effective masses of these quasiparticles are much less, than effective masses of "real" magnons – the quasiparticles describing small-scale exchange-interaction effects in magnetic structures [65]. The effective mass of the "light" magnon for a monochromatic MDM is defined as [81]:

$$\left(m_{lm}^{(eff)}\right)_n = \frac{\hbar}{2}\frac{\beta_n^2}{\omega}, \tag{21}$$

where $\beta_n$ is the propagation constant of mode *n* along the disk axis *z*.

In the *L*-mode solutions, we determine the membrane function $\tilde{\varphi}(r,\theta)$ at the same way as the membrane function $\tilde{\eta}(r,\theta)$ for *G*-mode solutions. The continuity of $\tilde{\varphi}(r,\theta)$ on a lateral surface of a ferrite disk is characterized by the equation like Eq. (19). However, for the derivatives on a lateral surface we have nonhomogeneous boundary conditions:

$$\mu\left(\frac{\partial \tilde{\varphi}_n}{\partial r}\right)_{r=\mathcal{R}^-} - \left(\frac{\partial \tilde{\varphi}_n}{\partial r}\right)_{r=\mathcal{R}^+} = -\left(i\mu_a \frac{1}{r}\frac{\partial \tilde{\varphi}_n}{\partial \theta}\right)_{r=\mathcal{R}^-}. \tag{22}$$

This is the EM boundary condition of continuity of the magnetic flux density on a lateral surface of a ferrite disk. In Eqs. (20) and (22), $\mu$ and $\mu_a$ are diagonal component and off-diagonal components of the permeability tensor $\ddot{\mu}$ [65]. When using the EM boundary conditions, it becomes obvious that the membrane function $\tilde{\varphi}(r,\theta)$ must not only be continuous and differentiable with respect to a normal to the lateral surface of the disk, but, because of the presence of a gyrotropy term, be also differentiable with respect to a tangent to this surface.

From Eq. (22), it follows that for a given direction of a bias magnetic field (which defines a sign of $\mu_a$), we observe both clockwise (CW) and counterclockwise (CCW) azimuthally propagating modes. In this case, it can be assumed that the strength of the scattering of light for the clockwise to the counterclockwise propagation direction is different. For homogeneous ND boundary conditions, we have Hermitian Hamiltonian resulting in real energy eigen states. However, since the sample is an open system, the coupling to the environment expressed by the EM boundary conditions leads to non-Hermitian Hamiltonian with complex wave functions. Membrane functions $\tilde{\varphi}(r,\theta)$ are not single-valued functions. It can be represented as a two-component spinor [85]:

$$\tilde{\varphi}_n(\vec{r},\theta) = \tilde{\eta}_n(\vec{r},\theta) \begin{bmatrix} e^{-\frac{1}{2}i\theta} \\ e^{+\frac{1}{2}i\theta} \end{bmatrix} \tag{23}$$

Circulation of gradient $\vec{\nabla}_\theta \tilde{\varphi}$ along contour $L = 2\pi r$ is not equal to zero. On a lateral border of a ferrite disk ($r = \mathcal{R}$), we express function $\tilde{\varphi}$ as

$$\tilde{\varphi} = \tilde{\eta}\delta_\pm, \tag{24}$$

where $\delta_\pm$ is a double-valued edge wave function on contour $\mathcal{L} = 2\pi\mathcal{R}$ [79, 82].

On a lateral surface of a quasi-2D ferrite disk, one can distinguish two different functions $\delta_\pm$, which are the counterclockwise and clockwise rotating-wave edge functions with respect to a membrane function $\tilde{\eta}(r,\theta)$. The spin-half wave-function $\delta_\pm$ changes its sign when the regular-coordinate angle $\theta$ is rotated by $2\pi$. As a result, one has the eigenstate spectrum of MDM oscillations with topological phases accumulated by the edge wave function $\delta$. A circulation of gradient $\vec{\nabla}_\theta \delta = \frac{1}{r}\left(\frac{\partial \delta_\pm}{\partial \theta}\right)_{r=\mathcal{R}} \vec{e}_\theta$ along contour $\mathcal{L} = 2\pi\mathcal{R}$ gives a non-zero quantity when an azimuth number is $q_\pm = \pm\frac{1}{2},\pm\frac{3}{2},\pm\frac{5}{2}...$ A line integral around a singular contour $\mathcal{L}$: $\frac{1}{\mathcal{R}}\oint_L \left(i\frac{\partial \delta_\pm}{\partial \theta}\right)(\delta_\pm)^* d\mathcal{L} = \int_0^{2\pi}\left[\left(i\frac{\partial \delta_\pm}{\partial \theta}\right)(\delta_\pm)^*\right]_{r=\mathcal{R}} d\theta$ is an observable quantity. It follows from the fact that because of such a quantity one can restore single-valuedness (and, therefore, Hermicity) of the spectral problem. We can represent this observable quantity as a linear integral of a certain vector potential. Because of the existing the geometrical phase factor on a lateral boundary of a ferrite disk, MDM oscillations are characterized by a pseudo-electric field (the gauge field) $\vec{\epsilon}$. The pseudo-electric field $\vec{\epsilon}$ can be found as $\vec{\epsilon}_\pm = -\vec{\nabla} \times \left(\vec{\Lambda}_\epsilon^{(m)}\right)_\pm$, where a vector function $\left(\vec{\Lambda}_\epsilon^{(m)}\right)_\pm$ can be considered as the Berry connection. The gauge-invariant field $\vec{\epsilon}$ is the Berry curvature. The corresponding flux of the field $\vec{\epsilon}$ through a circle of radius $\mathcal{R}$ is obtained as:, $K\int_S \left(\vec{\epsilon}\right)_\pm \cdot d\vec{S} = K\oint_\mathcal{L} \left(\vec{\Lambda}_\epsilon^{(m)}\right)_\pm \cdot d\vec{\mathcal{L}} = K\left(\Xi^{(e)}\right)_\pm = 2\pi q_\pm$, where $\left(\Xi^{(e)}\right)_\pm$ are quantized fluxes of pseudo-electric fields and $K$ is the normalization coefficient. The physical meaning of coefficient $K$ concerns the property of a flux of a pseudo-electric field. There are the positive and negative eigenfluxes. In the MDM spectral problem, it is impossible to satisfy the EM boundary conditions without a flux $\left(\Xi^{(e)}\right)_\pm$. Each MDM is characterized by energy eigenstate and is quantized to a quantum of an emergent electric flux.

In our system, there should be a certain internal mechanism which creates a nonzero vector potential $\left(\vec{\Lambda}_\epsilon^{(m)}\right)_\pm$. This internal mechanism becomes apparent when comparing the ND boundary condition (20) (providing single-valuedness) and the EM boundary condition (22) (which does not provide single-valuedness). The difference arises from the term in the right-hand side in Eq. (22), which contains the gyrotropy parameter, the off-diagonal component of the permeability tensor $\mu_a$, and the annular magnetic field $\vec{H}_\theta = -\frac{1}{r}\left(\frac{\partial \delta_\pm}{\partial \theta}\right)_{r=\mathcal{R}} \vec{e}_\theta$. Just due to this term a nonzero vector potential appears. The annular magnetic field $\vec{H}_\theta$ is a singular field existing only in an infinitesimally narrow cylindrical layer abutting from a ferrite side to a border of a ferrite disk. One does not have any special conditions connecting radial and azimuth components of magnetic fields on other inner or outer circular contours, except contour $\mathcal{L} = 2\pi\mathcal{R}$. Because of such an annular magnetic field, the notion of an effective circular magnetic current can be considered. The Berry mechanism provides a basis for the surface magnetic current at the interface between gyrotropic and nongyrotropic media. Following the spectrum analysis of MDMs in a quasi-2D ferrite disk one obtains edge chiral magnetic currents. This results in appearance of an anapole moment. For mode $n$, the anapole moment is calculated as [79, 82].

$$\left(a_\pm^{(e)}\right)_n \propto \mathcal{R}\int_0^d \oint_\mathcal{L} \left[\vec{j}_s^{(m)}(z)\right]_n \cdot d\vec{l}dz, \tag{25}$$

where $\vec{j}_s^{(m)}$ is the edge persistent magnetic current. At a large distance from the disk, localized distribution of an edge magnetic current is viewed as an electric field $\vec{a}^{(e)}$. The electric moment $\vec{a}^{(e)}$ is considered as the density of the electric flux $\left(\Xi^{(e)}\right)_\pm$.

Based on of Eq. (18) for the power-flow density, in Ref. [82] it was shown that the orthogonality conditions for the $L$-mode spectral solutions take place when

$$\int_0^{2\pi} \left(\mathcal{J}_\pm^{(s)}(z)\right)_\theta d\theta = \frac{1}{4}\omega\mu_0 \mathfrak{R} \int_0^{2\pi} \left[\left(i\mu_a \frac{\partial \delta_\pm}{\partial \theta}\right)^* \delta_\pm - \left(i\mu_a \frac{\partial \delta_\pm}{\partial \theta}\right)(\delta_\pm)^*\right]_{r=\mathfrak{R}} d\theta = 0 \qquad (26)$$

where $\left(\mathcal{J}_\pm^{(s)}\right)_\theta$ is a surface power-flow density on contour $\mathcal{L} = 2\pi\mathcal{R}$. It is evident, however, that the power-flow-confinement states can be realized when a softer boundary condition on contour $\mathcal{L}$ is used:

$$\vec{\nabla}_\theta \cdot \left[\left(\mathcal{J}_\pm^{(s)}(z)\right)_\theta\right]\vec{e}_\theta = \frac{1}{\mathfrak{R}} \frac{\partial}{\partial \theta}\left[\left(i\mu_a \frac{\partial \delta_\pm}{\partial \theta}\right)^* \delta_\pm - \left(i\mu_a \frac{\partial \delta_\pm}{\partial \theta}\right)(\delta_\pm)^*\right]_{r=\mathfrak{R}} = 0. \qquad (27)$$

This implies the presence of the edge persistent power-flow circulation.

Due to the surface power flow density, the membrane eigenfunction $\tilde{\eta}$ of the MDM rotates around the disk axis. When for every MDM we introduce the notion of an effective mass $\left(m_{lm}^{(eff)}\right)_n$, expressed by Eq. (21), we can assume that for every MDM there exists also an effective moment of inertia $\left(I_z^{(eff)}\right)_n$. With this assumption, an orbital angular momentum a mode is expressed as $(L_z)_n = \left(I_z^{(eff)}\right)_n \omega$. Supposing, as the first approximation, that the membrane eigenfunction $\tilde{\eta}_n$ is viewed as an infinitely thin homogenous disk of radius $\mathcal{R}$ (in other words, assuming that for every MDM, the radial and azimuth variation of the MS-potential function, are averaged), we can write for the effective moment of inertia

$$\left(I_z^{(eff)}\right)_n = \frac{1}{2}\left(m_{lm}^{(eff)}\right)_n \mathcal{R}^2 d. \qquad (28)$$

The orbital angular momentum a mode is expressed as

$$\left(L_z^{(eff)}\right)_n = \left(I_z^{(eff)}\right)_n \omega = \frac{\hbar}{4}\beta_n^2 \mathcal{R}^2 d. \qquad (29)$$

With use of the EM boundary conditions, we consider the spectral solutions for the MS wave functions $\psi$ as generating functions for determining the fields. For any mode $n$, magnetization field is found as $\vec{m} = -\frac{1}{4\pi}(\vec{\mu} - \vec{I})\vec{\nabla}\psi$ [65]. Knowing $\vec{\nabla}\cdot\vec{m}$ and $\vec{\nabla}\times\vec{m}$, we can obtain for the electric and magnetic fields outside a ferrite disk [79]:

$$\vec{E}(\vec{r}) = -\frac{i\omega\mu_0}{4\pi}\left(\int_V \frac{\vec{\nabla}'\times\vec{m}(r')}{|\vec{r}-\vec{r}'|}dV' + \int_S \frac{\vec{m}(r')\times\vec{n}'}{|\vec{r}-\vec{r}'|}dS'\right) \qquad (30)$$

and

$$\vec{H}(\vec{r}) = \frac{1}{4\pi} \left( \int_V \frac{\left(\vec{\nabla}' \cdot \vec{m}(r')\right)(\vec{r} - \vec{r}')}{\left|\vec{r} - \vec{r}'\right|^3} dV' - \int_S \frac{\left(\vec{n}' \cdot \vec{m}(r')\right)(\vec{r} - \vec{r}')}{\left|\vec{r} - \vec{r}'\right|^3} dS' \right), \tag{31}$$

where $V$ and $S$ are a volume and a surface of a ferrite sample, respectively. Vector $\vec{n}'$ is the outwardly directed normal to surface $S$.

In the vacuum near-field region adjacent to the MDM ferrite disk, there exist power-flow vortices, defined by the cross product $\text{Re}(\vec{E} \times \vec{H}^*)$. Together with this, there is another quadratic-form parameter determined by a scalar product between the electric and magnetic field components [79, 85]:

$$F = \frac{\varepsilon_0}{4} \text{Im}\left[\vec{E} \cdot \left(\nabla \times \vec{E}\right)^*\right] = \frac{\omega \varepsilon_0 \mu_0}{4} \text{Re}\left(\vec{E} \cdot \vec{H}^*\right) = \frac{\omega \varepsilon_0 \mu_0}{4} \text{Re}\left(\vec{\nabla}\phi \cdot \vec{\nabla}\psi^*\right) \tag{32}$$

The presence of this parameter, called the parameter of helicity, in the vacuum region is a fundamental effect in our analysis. In Ref. [86, 87], in was argued, that a linear structure of the EM radiation fields in vacuum can be observed only when $\vec{E} \cdot \vec{B} = 0$. At this condition, the electromagnetic helicity is defined as a difference between the numbers of right- and left-handed photons. On the other hand, when $\vec{E} \cdot \vec{B} \neq 0$, quantum electrodynamics predicts that the vacuum behaves like a material medium. In this case, the linear Maxwell theory receives nonlinear corrections. One can observe such a ME birefringence of the quantum vacuum when static magnetic and electric fields are applied [74, 88 – 90]. In our case, the nonlinearity in the vacuum near-field region adjacent to the MDM ferrite disk arises due to magnon-magnon dipole interaction effects. We observe the dominant magnonic response, even at room temperature. Large binding energy and small size of a MDM particle enables strong light-matter coupling to cavity photons and magnons, leading to emergent magnon-polaritons.

However, the question arises: What is the physics of quantization of magnetic dipole-dipole interactions in a quasi-2D ferrite disk? In Ref. [83], we have qualitatively explained how the multiresonance states in a microwave resonator, experimentally observed in [91, 92], are associated with a quantized change in the energy of a ferrite disk, which arises due to an external source – a bias magnetic field $H_0$ – at a constant frequency of the microwave signal. It was stated that there is a quantum effect of electromagnetically generated demagnetization of a sample:

$$\Delta W^{(n)} = -\frac{1}{2} \int_V \vec{H}_0^{(n)} \cdot \Delta\left(\vec{M}^{(n)}\right)_{eff} dV. \tag{34}$$

The energy $\Delta W^{(n)}$ is the microwave energy extracted from the magnetic energy of a ferrite disk at the $n$-th MDM resonance. It was supposed that the demagnetizing magnetic field is reduced due to effective magnetic charges on a ferrite-disk planes. For the reduced DC magnetization of a ferrite disk, we have the frequency $\left(\omega_M^{(n)}\right)_{eff} = \gamma\mu_0\left(M_0^{(n)}\right)_{eff}$, that is less than such a frequency $\omega_M = \gamma\mu_0 M_0$ in an unbounded magnetically saturated ferrite [65]. The quantized magnetic charges on the ferrite-disk planes are caused by the induced electric gyrotropy and orbitally driven electric polarization inside a ferrite. This is due to EDM oscillations in the electric subsystem.

**B. EDM oscillations in the electric subsystem**

The properties of dynamic magnetoelectricity are related to the processes of transition of magnetic energy into electric energy and vice versa. At the MDM resonances, the inhomogeneous magnetization structures are accompanied by the induced electric polarization properties. Since in a subwavelength ferrite-disk the EM retardation are neglected, the EDM oscillations should be described by the ES scalar wave functions. YIG is an electrically isotropic material. To solve analytically the ES-wave spectral problem in a quasi-2D ferrite disk, the effective parameters of the permittivity tensor of the induced electric anisotropy in should be found. When analyzing magnetically induced electrical anisotropy, we can initially use a simple model based on symmetry considerations. This is model of electric polarization induced by magnetic ordering of spiral spin waves [93]. It is evident that the invariance upon the time reversal, requires coupling between magnetization $\vec{m}$ and electric polarization $\vec{p}$ to be quadratic in $\vec{m}$. Also, the symmetry with respect to the spatial inversion, is respected when the coupling is linear in $\vec{p}$ and contains a gradient of $\vec{m}$. The electric polarization $\vec{p}$ can only couple to the magnetization if and only if $\vec{m}$ also has a directionality and lacks a center of inversion symmetry. In a spatially symmetric structure, the symmetry arguments give the following relationship for the density of ME energy [77, 93]:

$$W_{ME} \propto \vec{p} \cdot \left[ \vec{m}(\vec{\nabla} \cdot \vec{m}) - (\vec{m} \cdot \vec{\nabla})\vec{m} \right]. \tag{35}$$

The quantity $\vec{m}(\vec{\nabla} \cdot \vec{m}) - (\vec{m} \cdot \vec{\nabla})\vec{m}$ is called Lifshitz invariant. From the vector calculus identity $\left[ \vec{a}(\vec{\nabla} \cdot \vec{b}) - (\vec{a} \cdot \vec{\nabla})\vec{b} \right] = \vec{a} \times (\vec{\nabla} \times \vec{b})$, we obtain:

$$\vec{p} \propto \vec{m} \times (\vec{\nabla} \times \vec{m}). \tag{36}$$

We have the following components of the electric polarization:

$$p_x = \mathcal{K}\left[ \vec{m} \times (\vec{\nabla} \times \vec{m}) \right]_x = \mathcal{K} m_y \left( \frac{\partial m_y}{\partial x} - \frac{\partial m_x}{\partial y} \right),$$

$$p_y = \mathcal{K}\left[ \vec{m} \times (\vec{\nabla} \times \vec{m}) \right]_y = -\mathcal{K} m_x \left( \frac{\partial m_y}{\partial x} - \frac{\partial m_x}{\partial y} \right),$$

$$p_z = \mathcal{K}\left[ \vec{m} \times (\vec{\nabla} \times \vec{m}) \right]_z = \mathcal{K}\left( m_x \frac{\partial m_x}{\partial z} + m_y \frac{\partial m_y}{\partial z} \right), \tag{37}$$

where $\mathcal{K}$ is a coefficient of proportionality. In the ferrite-disk structure, a spiral spin-oscillation states appear due to the presence of the *z*-coordinate derivative of magnetization.

At the MDM resonances, polarization of the electric subsystem is due to electric flux $\Xi^{(e)}$. This flux polarizes the electric subsystem in accordance with general symmetry arguments regarding the form of coupling between the magnetization and electric polarization. The electric polarization in a ferrite disk is the rotation-field induced polarization. It occurs when the inversion symmetry of the magnetization-dynamics distribution is spontaneously broken. For a magnetically saturated ferrite disk with a normal bias magnetic field directed along the $z$ axis, the MDM magnetization distribution shows that the real-time electric polarization component $\vec{p}_z$ is an antisymmetric quantity when geometry of the microwave structure is symmetrical along $z$ axis [77]. Following this symmetry argument, we can see that the $z$-distribution of the currents $\vec{j}_s^{(m)}(z)$ in Eq. (25) is also antisymmetric. This situation is illustrated in Fig. 1.

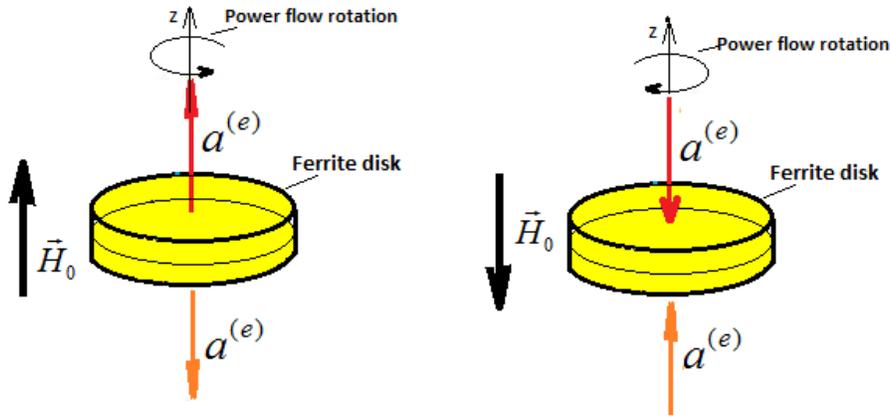

Fig. 1. Symmetric microwave structure. In the magnetic subsystem of the ferrite-disk resonator, localized distribution of an edge magnetic current is viewed as an electric flux. The red arrows indicate the z-direction of the electric flux density at the center of the disk. The electric moment $\vec{a}^{(e)}$ is considered as the density of the electric flux $\Xi^{(e)}$. We visualize quadrupole moment as two anti-parallel electric dipoles. Chiral magnetic currents at the lateral boundary induce a normal electric-field gradient. There is a structure with the opposite polarization directions on the top and bottom planes of a ferrite disk.

From Fig. 1, one can see that chiral magnetic currents at the lateral boundary induce a normal *electric-field gradient*. We visualize quadrupole moment as two anti-parallel electric dipoles. The electric-quadrupole precession is induced by the magnetization dynamics. Quadrupole moment leads to an asymmetric electric charge distribution of an electric subsystem. The dipole moment is zero, regardless of the coordinate origin that has been chosen. But the quadrupole moment cannot be reduced to zero, regardless of where we place the coordinate origin. In an electric field $\vec{a}^{(e)}$, each of electric dipoles of an electric subsystem experiences a torque tending to align them along the field; since the two turning torques are equal and opposite. However, in an electric field gradient which we assume to be axial-symmetric about the z axis, the torques on the two dipoles are not equal and a net turning torque exists which is clearly proportional both to the electric field gradient. The situation may be compared to the precessional model for magnetic dipoles when an applied magnetic field tries to turn the magnet into parallelism. The quadrupole interaction is particular interest in nuclear physics where they reflect the nature of the forces between neutrons and protons [94, 95]. In our case, the quadrupole interaction is related to the quasistatic interaction between the magnetic and electric subsystems in a ferrite-disk resonator. The quadrupole interaction is expressed as $\overset{\leftrightarrow}{Q} \otimes \vec{\nabla} \cdot \vec{a}^{(e)}$, where $\overset{\leftrightarrow}{Q}$ is the quadrupole moment tensor [74]. At the EDM resonances, the quadrupole moment tensor satisfies the eigenvalue equation.

For a given direction of a bias magnetic field, vectors of the electric-dipole moments $\vec{p}$ rotates synchronously with vectors of magnetization $\vec{m}$. This is shown in Fig. 2. We have the combined effects of magnetic dipole precession in a bias magnetic field along with electric quadrupole precession in an electric field gradient. It is worth noting that the frequency of orbital rotation is twice the frequency of precession [85]. In the laboratory coordinate system, we see a magnetic dipole and an electric quadrupole rotating at a frequency twice the frequency of the microwave signal (Fig. 3).

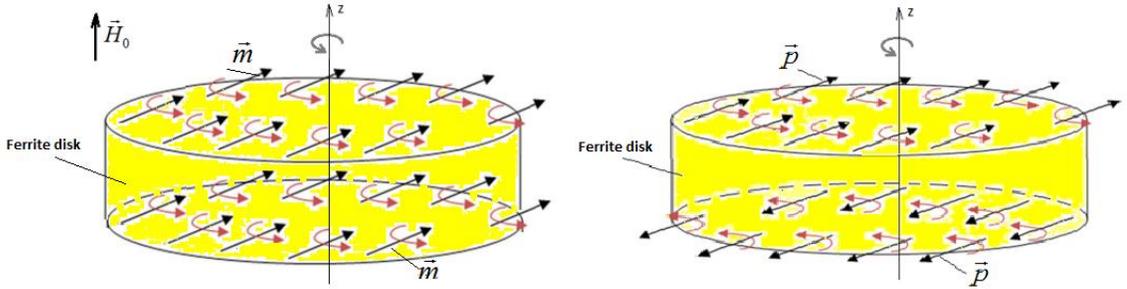

Fig. 2. For a given direction of a bias magnetic field, vectors of the electric-dipole moments $\vec{p}$ rotates synchronously with vectors of magnetization $\vec{m}$. The frequency of orbital rotation is twice the frequency of precession. We have the combined effects of magnetic dipole precession in a bias magnetic field along with electric quadrupole precession in an electric field gradient.

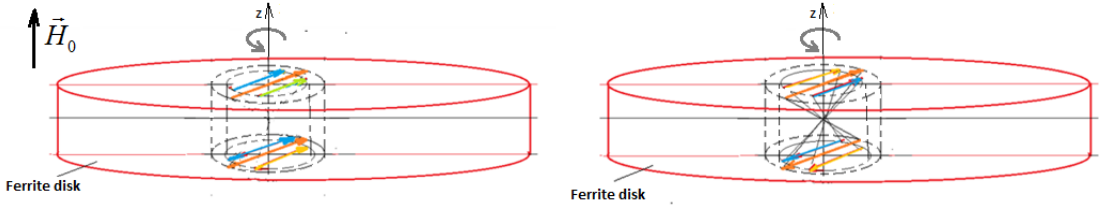

Fig. 3. In the laboratory coordinate system, we see a magnetic dipole and an electric quadrupole rotating at a frequency twice the frequency of the microwave signal.

When a ferrite disk is placed in a spatially non-symmetric microwave structure, the DC electric polarization of a disk can occur. Direction of the electric moment $\vec{a}^{(e)}$ is related to the direction of a bias magnetic field $\vec{H}_0$. This is shown in Fig. 4. The effect of electric disk polarization on MDM resonances in spatially asymmetric microwave structures was experimentally confirmed in Refs. [96 – 98]. It has also been used to probe dielectric samples. At dielectric loading of a ferrite sample, the Larmor frequency becomes lower than such a frequency for an unloaded ferrite disk. Due to loading by an external dielectric sample, effective DC magnetic charges appear on the planes of the ferrite disk. So, we have the ME effect of the appearance of both DC electric and magnetic charges on the planes of the ferrite disk at the MDM resonances. It is worth noting that in a case of a spatially non-symmetric microwave structure, we do not observe electric quadrupole precession. In Ref. [79] it was assumed that in this case, there should be a differential equation describing the precessional motion of electric dipoles, similar to the Landau–Lifshitz equation for the precessional motion of magnetic dipoles.

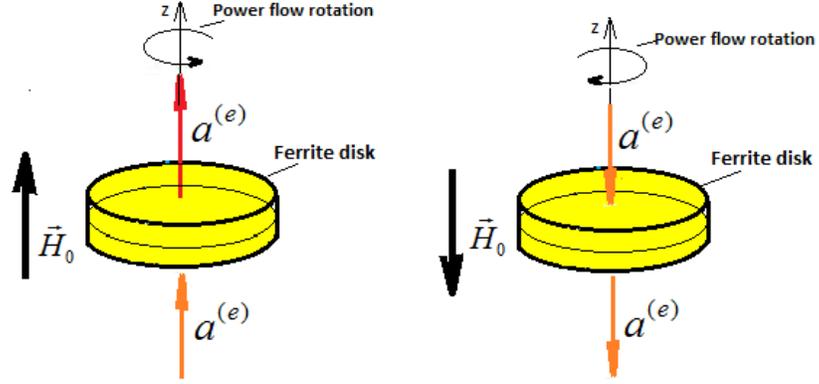

Fig. 4. The electric moment $\vec{a}^{(e)}$, when MDM ferrite is placed in an antisymmetric microwave structure.

The electric-dipole dynamics in a ferrite disk, induced at the MDM resonances, should be characterized by spectral eigenstates. Eigenfunctions of these electric dipolar oscillations are ES scalar wave functions $\phi$. The effective permittivity tensor $\vec{\bar{\varepsilon}}(\omega)$ of the induced electric anisotropy is the gyrotropic tensor. Parameters of this tensor can be found phenomenologically based on numerical and experimental analysis of the spectral characteristics. The spectral analysis for ES scalar wave functions $\phi$ in a quasi-2D ferrite disk will be like the spectral analysis for MS scalar wave functions $\psi$. Similarly to MDM oscillations, for EDM oscillations, the differences between the ND and EM boundary conditions will lead to the appearance of electric chiral currents at the lateral boundary of the disk sample. For both the above cases (when a ferrite disk is place in symmetric and antisymmetric microwave structures), we will have the physically observable magnetic flux $\Xi^{(m)}$. In the presence of double-valued functions in the EDM spectral problems, the fulfillment of the EM boundary conditions for the EDM without this flux is impossible. Localized distribution of an edge electric current is viewed as a *dipole magnetic field* at a large distance. This is shown in Fig. 5. Chiral electric currents at the lateral boundary induce normal DC magnetization. This effect of quantized magnetic charges on the ferrite-disk planes caused by EDM oscillations in the electric subsystem was discussed above.

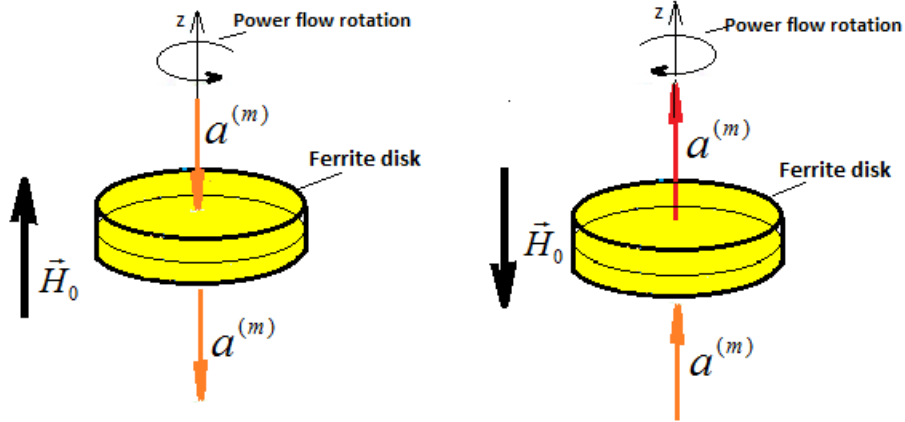

Fig. 5. Localized distribution of an edge electric current in the electric subsystem is viewed as a dipole magnetic field at a large distance. The arrows indicate the z-direction of the magnetic field flow at the center of the disk. The magnetic moment $\vec{a}^{(m)}$ is considered as the density of the magnetic flux $\Xi^{(m)}$.

## V. DISCUSSION: INTERACTION OF FERRITE ME META-ATOMS WITH EM FIELDS

The problem of EM wave scattering by proposed ME meta-atoms with coupled ES and MS oscillations seems to be rather complicated. The structure of the near fields of these subwavelength resonators – the ME fields – is defined by internal dynamical processes of magnetization and electric polarization in a quasi-2D ferrite disk and is distinguished by simultaneous violation of time reversal and inversion symmetry. It is evident that for small ferrite particles with quasistatic oscillations one cannot use the classical theory of Rayleigh scattering. Due to violation of a spatial symmetry of the fields, the standard method of expansion in terms of spherical or cylindrical harmonics is not applicable in this case. While for an incident wave there is no difference between left and right, in the fields scattered by a MDM ferrite particle one should distinguish left from right. To analyze the scattering problem, expansion of the fields by eigenmodes of quasistatic oscillations in a ferrite particle must be used. Generally, these oscillations are characterized by the helical-mode resonances.

In our structures, we have the orbital ME polarizability due to the "magnetic" and "electric" Berry connections. The ME field is a (3 + 1)-dimensional quantum field. It is well-known that while considering time as the fourth dimension, one is usually related to the second law of thermodynamics. There is, however, also the possibility to measure time by a so-called Larmor clock [99 – 101]. In our case, the spin-orbit interaction can be understood by considering a local magnetization vector with angular momentum orbiting around the disk axis. In the reference frame associated with the local magnetization vector, the disk rotates around it and creates an orbital magnetic current. The orbital magnetic current creates the electric field. In this field, precession of the magnetization vector records time of an orbital rotation. To calibrate the clock, we should correlate rotation of the magnetization vector with EM oscillations. We analyze the angle of rotation of the magnetization vector during the time of the RF period. In our structure, along with Larmor "magnetic" clocks, we should also talk about Larmor "electric" clocks. Both of these Larmor clocks are mutually synchronized.

While considering the ME field is a (3 + 1)-dimensional *PT* symmetric quantum field, we find an analogy with the axion-field electrodynamics and the axion-polariton problem. Definitely, in the ME-field electrodynamics we are talking about modified Maxwell equations with a presudoscalar source terms and the quantum field theory for massive particles. The $\vec{E}_{MDM} \cdot \vec{B}_{MDM}$ term of a MDM oscillation is the density of the ME energy. This is a pseudoscalar which couples nonlinearly to the external EM- field combination $\vec{E}_{EM} \cdot \vec{B}_{EM}$. However, when the axion fields are boson scalar fields, the ME fields are fermionic fields [83, 85].

Energy density of the ME near-field structure is defined by the helicity-factor density $F$. In a microwave structure when a ferrite disk is placed symmetrically with respect to metallic waveguide walls, a scalar parameter of helicity $F$, defined by Eq. (32), is asymmetric with respect to z axis. With that, the scalar parameter $F$ is asymmetric with respect to direction of a bias magnetic field. Distribution of the helicity factor in a vacuum region near a ferrite disk is shown in Fig. 6.

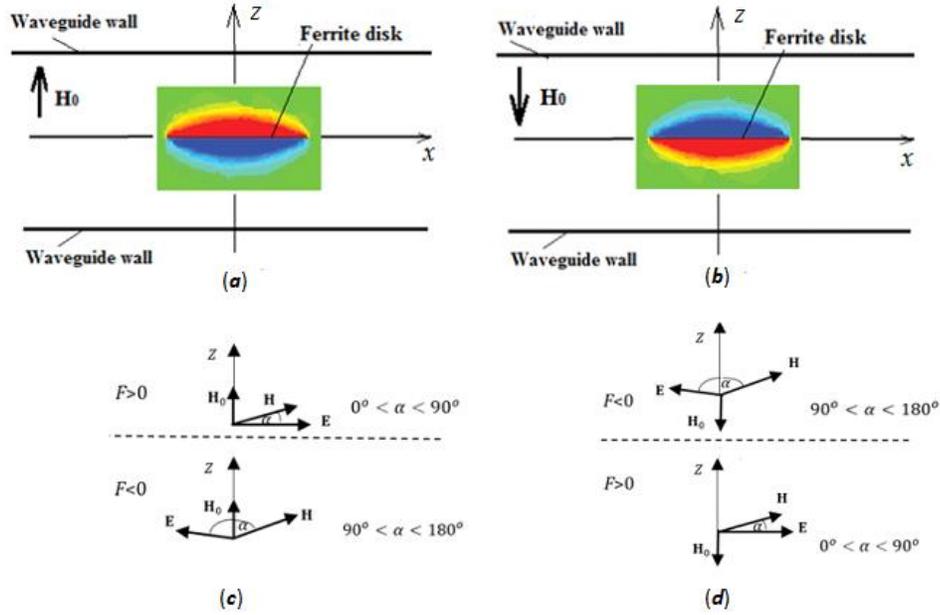

Fig. 6. The helicity factor $F$ distribution near a ferrite disk in a geometrically symmetrical microwave structure. The factor $F$ is characterized by antisymmetric distribution with respect to $z$ axis. Along with this, the helicity factor changes a sign at time reversal. (***a***) and (***b***) show helicity factor $F$ for two directions of a bias magnetic field. In a red region, $F > 0$, in a blue region, $F < 0$. A green region, $F = 0$, is an "electromagnetic background". For $F \neq 0$, mutual directions of the in-plane components of the electric and magnetic fields are shown in (***c***) and (***d***), in correspondence with (***a***) and (***b***).

In Fig. 7, the helicity factor $F$ distribution near a ferrite disk is shown in a geometrically non-symmetrical microwave structure. These are the pictures of four cases of the distributions when two external parameters – the disk position on $z$ axis and the direction of a bias magnetic field – change.

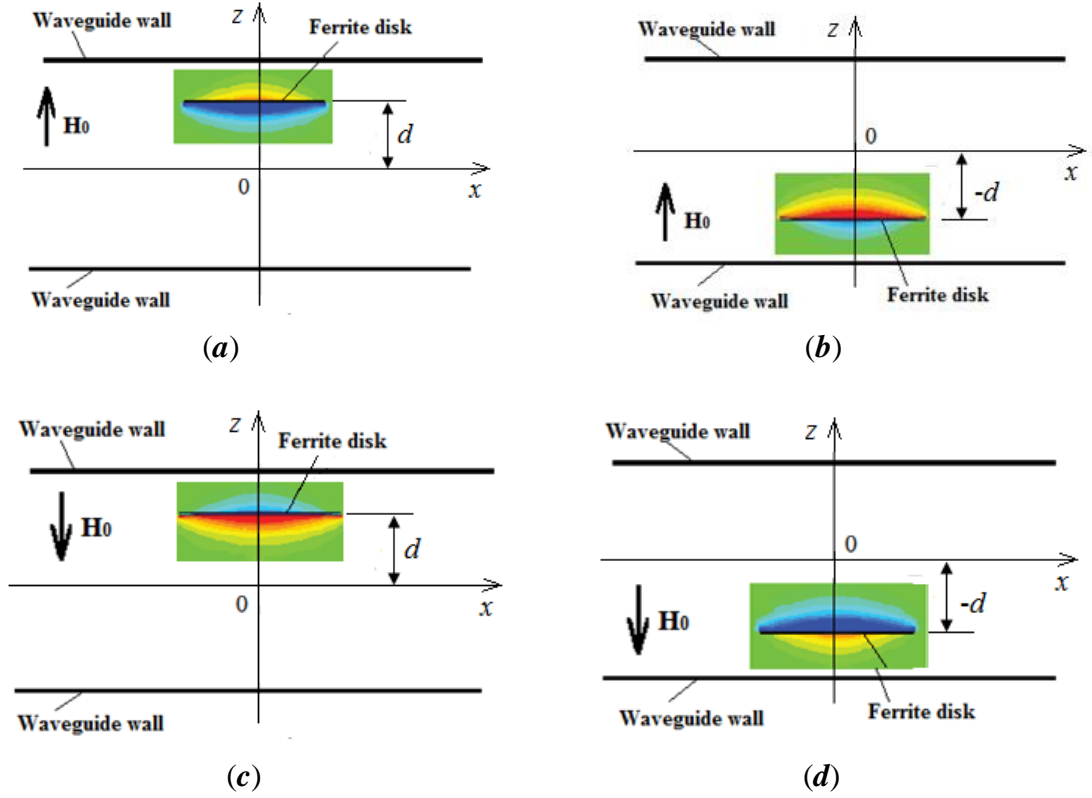

Fig. 7. The helicity factor $F$ distribution near a ferrite disk in a geometrically non-symmetrical microwave structure. Four cases of the helicity-factor distributions when two external parameters – the disk position on $z$ axis and the direction of a bias magnetic field – change. In a geometrically nonsymmetric structure, the distribution of a helicity factor $F$ becomes nonsymmetric as well. Reflection with respect to the $z = 0$ plane with simultaneous change of a direction of a bias magnetic field completely restore the same picture of a helicity factor $F$ [look at the distributions (*a*) and (*d*) and the distributions (*b*) and (*c*)]. At such a nonsymmetric disk position, the *PT* symmetry of MDM oscillations is broken and the biorthogonality condition is not satisfied.

In the shown microwave structures, we have the regions with the positive and negative ME-field densities, which are characterized by the positive and negative helicity-factor densities, $F^{(+)}$ and $F^{(-)}$, respectively. We define the positive helicity (the positive ME energy) as an integral of the ME-field density over the entire near-field vacuum region $\mathbb{V}^{(+)}$ with the helicity-factor density $F^{(+)}$:

$$\mathbb{H}^{(+)} = \int_{\mathbb{V}^{(+)}} F^{(+)} d\mathbb{V} \tag{38}$$

Similarly, the negative helicity (the negative ME energy) is defined as an integral of the ME-field density over the entire near-field vacuum region $\mathbb{V}^{(-)}$ with the helicity-factor density $F^{(-)}$:

$$\mathbb{H}^{(-)} = \int_{\mathbb{V}^{(-)}} F^{(-)} d\mathbb{V} . \tag{39}$$

In a case of *PT* symmetrical ME near-field distribution, shown in Fig. 6, there should be $\mathbb{H}^{(+)} = -\mathbb{H}^{(-)}$. So, the total helicity (the total ME energy) of the entire near-field vacuum region surrounding a ferrite disk $\mathbb{V} = \mathbb{V}^{(+)} + \mathbb{V}^{(-)}$ is equal to zero:

$$\mathbb{H} = \mathbb{H}^{(+)} + \mathbb{H}^{(-)} = \int_{\mathbb{V}} \left( F^{(+)} + F^{(-)} \right) d\mathbb{V} = 0. \tag{40}$$

If, however, the *PT* symmetry is broken (in a case shown, for example, in Fig. 7 when a ferrite disk is shifted in a waveguide along $z$ axis), we have $\left|\mathbb{H}^{(+)}\right| \neq \left|\mathbb{H}^{(-)}\right|$. It means that we may have predominance of the positive or negative ME energy. Figs. 7 (***a, d***) we have predominance of the negative ME energy, while in Figs. 7 (***b, c***), there is predominance of the positive ME energy. On a metal wall, the helicity-factor density (and so, the ME-energy density) is zero. Where does the excess of ME energy go? There is a chiral non-symmetric microwave structure. A waveguide section with non-symmetric positions of a ferrite disk, shown in Fig. 7, will be a non-reciprocal microwave structure with non-unitary scattering matrix.

Returning to a geometrically symmetrical structure, we consider a microwave waveguide structure in Fig.6 restricted with the two *xy*-plane metallic walls. In such a cavity, excitation of MDM oscillations in a quasi-2D ferrite disk occurs in a region where a normal component of the RF electric field is zero [78, 91, 92, 102]. Typical spectral characteristics are shown in Fig. 8.

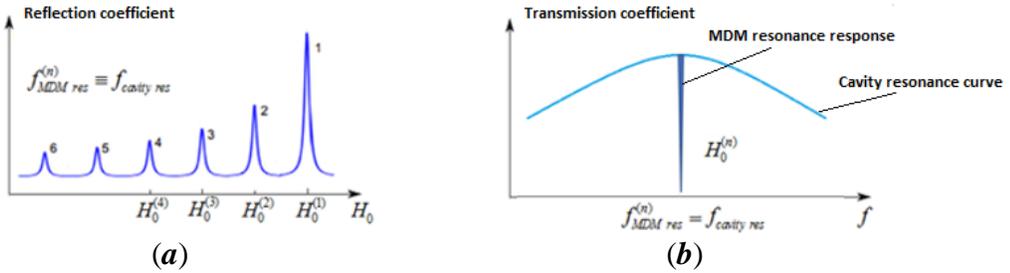

(***a***)                  (***b***)

Fig. 8. Typical spectral characteristics of MDM resonances in a microwave cavity. (***a***) The multiresonance spectrum at a variation of a bias magnetic field and a constant frequency. (***b***) A single Lorentzian peak of a MDM resonance at the cavity resonance frequency.

The multiresonance spectrum is observed at a variation of a bias magnetic field, provided that the frequency of each MDM resonance is equal to the cavity resonance frequency. The scattering cross section of a single Lorentzian peak at the state $f_{MDM\ res}^{(n)} = f_{cavity\ res}$ and a certain $H_0 = H_0^{(n)}$ is a pure dark mode. In an analysis of a field structure at dark-mode singular points, two models can be suggested.

(***A***) As it is argued in Ref. [85], the dark mode state is the lower energy state of the internal energy of the ferrite disk without the orbital rotation of MDM oscillation. With such an assumption, at the singular point $f_{MDM\ res}^{(n)} = f_{cavity\ res}$, no power-flow vortex is observed and the time by Larmor clock stops. To understand why the time by Larmor clock can stop at dark mode, the following explanation is used. Because of the electric field originated from a ferrite disk at the MDM resonance, every separate electric dipole in a disk precesses around its own axis. For all the precessing dipoles, there is an orbital phase running. A torque exerting on the electric polarization in a dielectric subsystem should be equal to a reaction torque exerting on the magnetization in a magnetic subsystem. In virtue of this reaction torque, the precessing magnetic moment density of the ferromagnet will be under additional mechanical rotation. In a dark-mode singular point the time by Larmor clock stops because the power-flow vortex of the magnetic subsystem is suppressed by the power-flow vortex of dielectric subsystem. As a result, we can have a quantized effect of ME coupling where both magnetic and electric dipoles precess but without orbital rotations.

(**B**) A quasi-2D ferrite disk in a microwave cavity is a non-Hermitian system. At the MDM resonance, it can support entirely real eigenvalue spectra in a so-called exceptional point (EP) at which two or more eigenvalues and underlying eigenvectors coalesce. EPs are associated with sharp phase transitions in the eigenvalue spectrum that can dramatically alter the response of the system. When we admit that the EP exists, we should observe a coalescence of the clockwise and counterclockwise azimuthal modes. Since the term on the right-hand side of Eq. (22) has only the first order derivative with respect to the azimuthal coordinate, such a coalescence does not lead to the presence of an azimuthal standing wave. The EP can be formed when the counter propagating azimuthal modes of a disk resonator coalesce into one traveling mode. In other words, the degenerate eigenmode has a preferred azimuth direction. As an example of such a special class of EPs, called chiral EPs (CEPs), we can refer to a structure of whispering gallery modes in optical microdisks [103, 104]. In our case of a quasi-2D ferrite disk, CEPs arise due to specific properties of the magnetic dipole-dipole dynamics. Since, herewith, the coupling with the environment means the interaction of a ferrite sample with an external microwave structure, one should assume that there is also a chiral structure of the vacuum field at the CEPs. This is due to the conditions of conservation of the angular momentum of the system when light is scattered on a ferrite disk. We observe photons with curved wave fronts of circularly polarized waves. These are a structure of twisted polaritons (See Fig. 9).

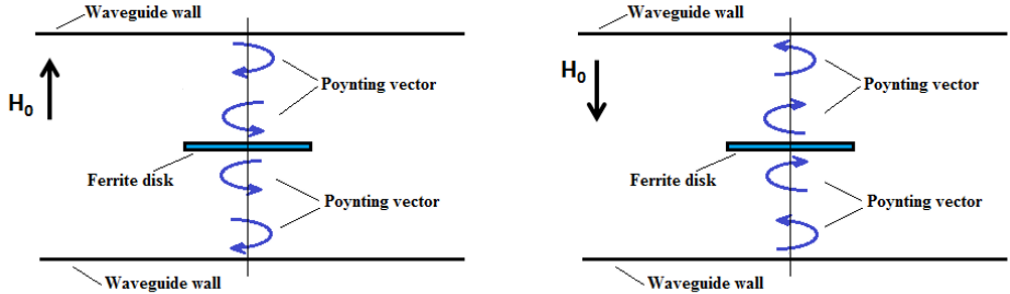

Fig. 9. Evidence for curved time-space in a microwave cavity. Angular momentum conservation at a chiral EPs (CEP) point. MDM oscillations are generated and modified due to EM radiation. On the other hand, due to these topological states, one can modify the EM radiation continuum. When the angular momentum balance is destroyed, we have the angular momentum transfer.

It is with noting that in a dark-mode singular point, effective interaction of ME meta-atoms with EM fields can be observed even in loss-dominated microwave systems. An innovative sensing technique for different types of high lossy materials placed inside a microwave cavity, shown in Refs. [105, 106], gives an example of a structure with so-called ultrastrong coupling regime.

## VI. CONCUSSION

Connection between magnetoelectricity and electromagnetism is a subject of great interest and numerous discussions in microwave and optical wave physics and material sciences. Like EM phenomena, described by Maxwell equations, the physics of ME phenomena deals with the problems of the relationship between electric and magnetic fields. Fundamental questions related to the restrictions imposed by the space dimension, symmetry, and the energy conservation law on the type and properties of the singular points – 3D confined ME resonant regions – remain open. The ME near field of a subwavelength ME particle can be characterized by space-time symmetry, which is different from space-time symmetry of the EM field. Free-space coupling to strongly subwavelength ME resonant elements arises as one of the central themes in quantum electrodynamics.

In this paper we showed that the realization of point ME scatterer is possible when we have a subwavelength resonator with both space and time breaking effects. In a quasi-2D ferrite disk, ME excitations are considered as combined excitations of the magnetization and electric-polarization orders. The ME interaction is expressed by coupling of the MS and ES oscillations. We clarify the concept of ME energy density. The near-fields of open quasistatic ME resonators are topological-nature fields – the ME fields. Our purpose is to study ME fields and EM problems associated with these fields. In a case of ME-field electrodynamics, we have to consider the Dirac equation (instead of the Klein–Gordon equation). That is, we have the equation for massive spin 1/2 particles. Together with the axion mass we have to consider also the axion moment of inertia. While the axion fields are boson scalar fields, the ME fields are fermionic fields. These fields are characterized by energy eigenstates with rotational superflows and quantized vortices.

In this connection, we discussed a problem of "bianisotropic meta-atoms". In a case of bianisotropic metamaterials, we have just a far-field view of magnetoelectricity. Due to non-acquaintance of the near-field magnetoelectricity of local 3D confined subwavelength structure of "bianisotropic meta-atoms", the question on ME properties of such point scatterers remains open. We can foresee numerous effects associated with the very idea of bianisotropy. We can also have numerous experimental tests of these models. But until we can determine the properties of the ME near fields, we will mislead ourselves about the basic physics of the bianistropic metamaterial structures.

**References**

[1] M. Kadic, G. W. Milton, M. van Hecke, and M. Wegener, "3D metamaterials, Nat. Rev. Phys. **1**, 198, (2019).

[2] I. Sersic, C. Tuambilangana, T. Kampfrath, and A. F. Koenderink, "Magnetoelectric point scattering theory for metamaterial scatterers", Phys. Rev. B **83**, 245102 (2011).

[3] C. E. Kriegler, M. S. Rill, S. Linden, and M. Wegener, "Bianisotropic photonic metamaterials", IEEE J. Select. Topics Quant. Electron. **16**, 367 (2010).

[4] A. Canaguier-Durand, J. A. Hutchison, C. Genet, and T. W Ebbesen, "Mechanical separation of chiral dipoles by chiral light", New J. Phys. **15**, 123037 (2013).

[5] M. Albooyeh, V. S. Asadchy, R. Alaee, S. M. Hashemi, M. Yazdi, M. S. Mirmoosa, C. Rockstuhl, C. R. Simovski, and S. A. Tretyakov, "Purely bianisotropic scatterers", Phys. Rev. B **94**, 245428 (2016).

[6] J. Proust, N. Bonod, J. Grand, and B. Gallas, "Optical monitoring of the magnetoelectric coupling in individual plasmonic scatterers", ACS Photonics **3**, 1581 (2016).

[7] V. S. Asadchy and S. A. Tretyakov, "Modular analysis of arbitrary dipolar scatterers", Phys. Rev. Applied **12**, 024059 (2019).

[8] C. Caloz and A. Sihvola, "Electromagnetic chirality", arXiv:1903.09087.

[9] M. Norgen, "General optimization approach to a frequency-domain inverse problem of a stratified bianisotropic slab", J. Electromagn. Waves Appl. **11**, 515 (1997).
[10] D. Sheen and D. Shepelsky, "Uniqueness in a frequency-domain inverse problem of a stratified uniaxial bianisotropic medium", Wave Motion, **31**, 371 (2000).
[11] X. Chen, T. M. Grzegorczyk, B.-I. Wu, J. Pacheco, Jr., and J. A. Kong, "Robust method to retrieve the constitutive effective parameters of metamaterials", Phys. Rev. E **70**, 016608 (2004).
[12] Z. F. Li, K. Aydin, E. Ozbay, "Determination of the effective constitutive parameters of bianisotropic metamaterials from reflection and transmission coefficients", Phys. Rev. E **79**, 026610 (2009).
[13] Y. Tang and A. E. Cohen, "Optical chirality and its interaction with matter," Phys. Rev. Lett. **104**, 163901 (2010).
[14] E. Hendry, T. Carpy, J. Johnston et al., "Ultrasensitive detection and characterization of biomolecules using superchiral fields," Nat. Nanotechnol. **5**, 783 (2010).
[15] M. M. Coles and D. L. Andrews, "Chirality and angular momentum in optical radiation," Phys. Rev. A **85**, 063810 (2012).
[16] K. Y. Bliokh, A. Y. Bekshaev, and F. Nori, "Dual electromagnetism: Helicity, spin, momentum and angular momentum," New J. Phys. **15**, 033026 (2013).
[17] A. Canaguier-Durand, J. A. Hutchison, C. Genet, and T. W. Ebbesen, "Mechanical separation of chiral dipoles by chiral light," New J. Phys. **15**, 123037 (2013).
[18] R. P. Cameron, S. M. Barnett, and A. M. Yao, "Discriminatory optical force for chiral molecules," New J. Phys. **16**, 013020 (2014).
[19] A. Canaguier-Durand and C. Genet, "Chiral route to pulling optical forces and left-handed optical torques," Phys. Rev. A **92**, 043823 (2015).
[20] E. Mohammadi, K. L. Tsakmakidis, A. N. Askarpour, P. Dehkhoda, A. Tavakoli, and H. Altug, "Nanophotonic platforms for enhanced chiral sensing", ACS Photonics **5**, 2669 (2018).
[21] L. E. Barr, S. A. R. Horsley, I. R. Hooper, J. K. Eager, C. P. Gallagher, S. M. Hornett, A. P. Hibbins, and E. Hendry, "Investigating the nature of chiral near-field interactions", Phys. Rev. B **97**, 155418 (2018).
[22] S. Droulias and L. Bougas, "Surface plasmon platform for angle-resolved chiral sensing", ACS Photonics **6**, 1485 (2019).
[23] T. V. Raziman, R. H. Godiksen, M. A. Müller, and A. G. Curto, "Conditions for enhancing chiral nanophotonics near achiral nanoparticles", ACS Photonics **6**, 2583 (2019).
[24] C. Hao, L. Xu, H. Kuang, and C. Xu, "Artificial chiral probes and bioapplications", Adv. Mater., **32**, 1802075 (2020).
[25] J. Mun, M. Kim, Y. Yang, T. Badloe, J. Ni, Y. Chen, C.-W. Qiu, and J. Rho, "Electromagnetic chirality: from fundamentals to nontraditional chiroptical phenomena", Light Sci. Appl. **9**, 139 (2020).
[26] S. Both, M. Schäferling, F. Sterl, E. Muljarov, H. Giessen, and T. Weiss, "Nanophotonic chiral sensing: How does it actually work?", ACS Nano **16**, 2822 (2022).
[27] D. Ayuso, O. Neufeld, A. F. Ordonez, P. Decleva, G. Lerner, O. Cohen, M. Ivanov and O. Smirnova, "Synthetic chiral light for efficient control of chiral light–matter interaction", Nature Photon. **13**, 866 (2019).
[28] O. Neufeld, M. E. Tzur, and O. Cohen, "Degree of chirality of electromagnetic fields and maximally chiral light", Phys. Rev. A **101**, 053831 (2020).
[29] O. Neufeld and O. Cohen, "Unambiguous definition of handedness for locally-chiral light", Phys. Rev. A **105**, 023514 (2022).
28

[30] K. Y. Bliokh, Y. S. Kivshar, and F. Nori, "Magnetoelectric effects in local light-matter interactions", Phys. Rev. Lett. **113**, 033601 (2014).
[31] M. Nieto-Vesperinas, "Optical theorem for the conservation of electromagnetic helicity: Significance for molecular energy transfer and enantiomeric discrimination by circular dichroism", Phys. Rev. A **92**, 023813 (2015).
[32] W. L. Barnes, A. Dereux, and T. W. Ebbesen, "Surface plasmon subwavelength optics", Nature **424**, 824 (2003).
[33] H. Katsura, N. Nagaosa, and A. V. Balatsky, "Spin current and magnetoelectric effect in noncollinear magnets", Phys. Rev. Lett. **95**, 057205 (2005).
[34] Y. Takahashi, R. Shimano, Y. Kaneko, H. Murakawa, and Y. Tokura, "Magnetoelectric resonance with electromagnons in a perovskite helimagnet", Nature Phys **8**, 121 (2012).
[35] Y. Takahashi, Y. Yamasaki, and Y. Tokura, "Terahertz magnetoelectric resonance enhanced by mutual coupling of electromagnons", Phys. Rev. Lett. **111**, 037204 (2013).
[36] S. Iguchi, R. Masuda, S. Seki, Y. Tokura, and Y. Takahashi, "Enhanced gyrotropic birefringence and natural optical activity on electromagnon resonance in a helimagnet", Nat. Commun. **12**, 6674 (2021).
[37] D. Szaller, A. Shuvaev, A. A. Mukhin, A. M. Kuzmenko, and A. Pimenov, "Controlling of light with electromagnons", Phys. Sci. Rev. **5**, 20190055 (2019).
[38] D. L. Mills, "Polaritons: the electromagnetic modes of media", Rep. Prog. Phys. **37**, 817 (1974).
[39] D. N. Basov, A. Asenjo-Garcia, P. J. Schuck, X. Zhu, and A. Rubio, "Polariton panorama", Nanophoton. **10**, 549 (2021).
[40] V. Gunawan and R. L. Stamps,"Surface and bulk polaritons in a PML-type magnetoelectric multiferroic with canted spins: TE and TM polarization", J. Phys.: Condens. Matter **23**, 105901 (2011).
[41] V. Gunawan and H. Widiyandari, "Polaritons in magnetoelectric multiferroics films: Switching magnon polaritons using an electric field", Ferroelectrics, **510**, 16 (2017).
[42] C. Weisbuch, M. Nishioka, A. Ishikawa, and Y. Arakawa, "Observation of the coupled exciton-photon mode splitting in a semiconductor quantum microcavity," Phys. Rev. Lett. **69**, 3314, (1992).
[43] J. Kasprzak, M. Richard, S. Kundermann, et al., "Bose–Einstein condensation of exciton polaritons" Nature. **443**, 409 (2006).
[44] I. Carusotto and C. Ciuti, "Quantum fluids of light," Rev. Mod. Phys. **85**, 29 (2013).
[45] T. Byrnes, N. Y. Kim, and Y. Yamamoto, "Exciton-polariton condensates," Nat. Phys. **10**,.803 (2014).
[46] A. Griffin, D. W. Snoke, and S. Stringari, eds., *Bose–Einstein Condensation* (Cambridge University Press, 1996).
[47] A. A. High, J. R. Leonard, A. T. Hammack, et al., "Spontaneous coherence in a cold exciton gas," Nature **483**, 7391 (2012).
[48] D. Snoke, "Spontaneous Bose coherence of excitons and polaritons," Science **298**, 1368 (2002).
[49] J. P. Eisenstein and A. H. MacDonald, "Bose–Einstein condensation of excitons in bilayer electron systems," Nature **432**, 691 (2004).
[50] H. Deng, H. Haug, and Y. Yamamoto, "Exciton-polariton Bose Einstein condensation," Rev. Mod. Phys. **82**, 1489 (2010).
[51] D. W. Snoke and J. Keeling, "The new era of polariton condensates," Phys. Today **70**, 54, (2017).
[52] Y. Sun, P. Wen, Y. Yoon, et al., "Bose–Einstein condensation of long-lifetime polaritons in thermal equilibrium," Phys. Rev. Lett. **118**, 016602 (2017).

[53] F. Wilczek, "Two applications of axion electrodynamics", Phys. Rev. Lett. **58**, 1799 (1987).
[54] A. M. Essin, J. E. Moore, and D. Vanderbilt, "Magnetoelectric polarizability and axion electrodynamics in crystalline insulators", Phys. Rev. Lett. **102**, 146805 (2009).
[55] R. Li, J. Wang, X.-L. Qi, and S.-C. Zhang, "Dynamical axion field in topological magnetic insulators", Nature Phys. **6**, 284 (2010).
[56] T. Ochiai, "Theory of light scattering in axion electrodynamics", J. Phys. Soc. Japan **81**, 094401 (2012).
[57] D. M. Nenno, C. A. C. Garcia, J. Gooth, C. Felser, and P. Narang, "Axion physics in condensed-matter systems". *Nat Rev Phys* **2,** 682–696 (2020).
[58] A. Sekine and K. Nomura, "Axion electrodynamics in topological materials", J. Appl. Phys. **129**, 141101 (2021).
[59] Y. Xiao, H. Wang, D. Wang, R. Lu, X. Yan, H. Guo, C.-M. Hu, K. Xia, H. Zhang, and D. Xing, "Nonlinear level attraction of cavity axion polariton in antiferromagnetic topological insulator", Phys. Rev. B **104**, 115147 (2021).
[60] J. Planelles, "Axion electrodynamics in topological insulators for beginners". arXiv:2111.07290 (2021).
[61] F. W. Hehl, Y. N. Obukhov, J.-P. Rivera, and H. Schmid, "Relativistic nature of a magnetoelectric modulus of $Cr_2O_3$ crystals: A four-dimensional pseudoscalar and its measurement", Phys. Rev. A **77**, 022106 (2008).
[62] F. W. Hehl, Y. N. Obukhov, J.-P. Rivera, and H. Schmid, "Magnetoelectric $Cr_2O_3$ and relativity theory", Eur. Phys. J. B **71**, 321 (2009).
[63] K. T. McDonald, "Magnetostatic spin waves", arXiv:physics/0312026, 2003
[64] L. R. Walker, "Magnetostatic modes in ferromagnetic resonance", Phys. Rev. **105**, 390 (1957).
[65] A. G. Gurevich and, G. A. Melkov, *Magnetization oscillations and waves* (CRC Press: New York, 1996).
[66] D. D. Stancil, *Theory of Magnetostatic Waves* (Springer Verlag: New York, 1993).
[67] L. D. Landau and E. M. Lifshitz, *Electrodynamics of Continuous Media* (Pergamon Press: Oxford, 1960).
[68] D. Mattis, *The Theory of Magnetism* (Harper & Row Publ.: New York, 1965).
[69] A. I. Akhiezer, V. G. Bar'yakhtar, and S. V. Peletminskii, *Spin Waves* (North-Holland: Amsterdam, 1968).
[70] K. T. McDonald, "An electrostatic wave", arXiv:physics/0312025, 2003.
[71] M. Kharitonov, "Interaction-enhanced magnetically ordered insulating state at the edge of a two-dimensional topological insulator", Phys. Rev. B **86**, 165121 (2012).
[72] X.-L. Qi and S.-C. Zhang, "Topological insulators and superconductors", Rev. Mod. Phys. **83**, 1057 (2011).
[73] M. Lang, M. Montazeri, M. C. Onbasli, X. Kou et al, "Proximity induced high-temperature magnetic order in topological insulator - ferrimagnetic insulator heterostructure", Nano Lett. **14**, 3459 (2014).
[74] J. D. Jackson, *Classical Electrodynamics*, 2nd ed. (Wiley, New York, 1975).
[75] P. M. Morse and H. Feshbach, *Methods of Theoretical Physics* (McGraw-Hill, New York, 1953).
[76] M. Sigalov, E. O. Kamenetskii, and R Shavit "Magnetic-dipolar and electromagnetic vortices in quasi-2D ferrite discs", J. Phys.: Condens. Matter **21**, 016003 (2009).
[77] E. O. Kamenetskii, M. Sigalov, and R. Shavit, "Tellegen particles and magnetoelectric metamaterials", J. Appl. Phys. **105**, 013537(2009).

[78] E. O. Kamenetskii, M. Sigalov, and R. Shavit, "Manipulating microwaves with magnetic-dipolar-mode vortices", Phys. Rev. A **81**, 053823 (2010)

[79] E. O. Kamenetskii, R. Joffe, and R. Shavit, "Microwave magnetoelectric fields and their role in the matter-field interaction," Phys. Rev. E **87**, 023201 (2013).

[80] E. O. Kamenetskii, "Energy eigenstates of magnetostatic waves and oscillations", Phys. Rev. E **63**, 066612 (2001).

[81] E. O. Kamenetskii, R. Shavit, and M. Sigalov. "Quantum wells based on magnetic-dipolar-mode oscillations in disk ferromagnetic particles", Europhys. Lett. **64**, 730 (2003).

[82] E. O. Kamenetskii, "Vortices and chirality of magnetostatic modes in quasi-2D ferrite disc particles", J. Phys. A: Math. Theor. **40**, 6539 (2007).

[83] E. O. Kamenetskii, "Quantization of magnetoelectric fields", J. Mod. Opt. **66**, 909 (2019).

[84] E. O. Kamenetskii, "Quasistatic oscillations in subwavelength particles: can one observe energy eigenstates?", Ann. Phys. (Berlin) **531**, 1800496 (2019).

[85] E. O. Kamenetskii, "Magnetic dipolar modes in magnon-polariton condensates", J. Mod. Opt. **68**, 1147 (2021).

[86] A. F. Ranada, "Topological electromagnetism", J. Phys.: Math. Gen. **25**, 1621 (1992).

[87] J. L. Trueba and A. F. Ranada, "The electromagnetic helicity", Eur. J. Phys. **17**, 141 (1996).

[88] G. L. J. A. Rikken and C. Rizzo, "Magnetoelectric anisotropy of the quantum vacuum", Phys. Rev. A **67**, 015801 (2003).

[89] R. Battesti and C. Rizzo, "Magnetic and electric properties of a quantum vacuum", Rep. Prog. Phys. **76**, 016401 (2013).

[90] A. D. Bermúdez Manjarres and M. Nowakowski, "Travelling waves in the Euler-Heisenberg electrodynamics", Phys. Rev. A **95**, 043820 (2017).

[91] J. F. Dillon Jr., "Magnetostatic modes in disks and rods", J. Appl. Phys. **31**, 1605 (1960).

[92] T. Yukawa and K. Abe, "FMR spectrum of magnetostatic waves in a normally magnetized YIG disk", J. Appl. Phys. **45**, 3146 (1974).

[93] M. Mostovoy, "Ferroelectricity in spiral magnets", Phys. Rev. Lett. **96**, 067601 (2006).

[94] M. Bloom, E. L. Hahn, and B. Herzog, "Free magnetic induction in nuclear quadrupole resonance", Phys. Rev. **97**, 1699 (1955).

[95] J. A. S. Smith, "Nuclear quadrupole resonance spectroscopy. General principles", J. Chem. Educ. **48**, 1, 39 (1971).

[96] E. O. Kamenetskii, A. K. Saha, and I. Awai, "Interaction of magnetic-dipolar modes with microwave-cavity electromagnetic fields", Physics Letters A **332**, 303 (2004).

[97] M. Sigalov, E. O. Kamenetskii, and R. Shavit, "Eigen electric moments and magnetic–dipolar vortices in quasi-2D ferrite disks", Appl. Phys. B **93**, 339 (2008).

[98] M. Sigalov, E. O. Kamenetskii, and R. Shavit "Electric self-inductance of quasi-two-dimensional magnetic-dipolar-mode ferrite disks", J. Appl. Phys. **104**, 053901 (2008).

[99] J. Kaushal, F. Morales, L. Torlina, M. Ivanov and O. Smirnova, "Spin–orbit Larmor clock for ionization times in one photon and strong-field regimes", J. Phys. B: At. Mol. Opt. Phys. **48**, 234002 (2015).

[100] D. Sokolovski and E. Akhmatskaya, "Tunnelling times, Larmor clock, and the elephant in the room", Sci. Rep. **11**, 10040 (2021).

[101] C. Zhao, X. Ma, H. Huang, et al. "Micromagnetic simulation of electric field modulation on precession dynamics of spin torque nano-oscillator", Appl. Phys. Lett. **111**, 082406 (2017).

[102] M. Berezin, E. O. Kamenetskii, and R. Shavit, "Topological-phase effects and path-dependent interference in microwave structures with magnetic-dipolar-mode ferrite particles", J. Opt. **14**, 125602 (2012).

[103] J. Wiersig, "Structure of whispering-gallery modes in optical microdisks perturbed by nanoparticles", Phys. Rev. A **84**, 063828 (2011).

[104] A. Hashemi, S. M. Rezaei, S. K. Özdemir, and R. El-Ganainy, "New perspective on chiral exceptional points with application to discrete photonics", APL Photon. **6**, 040803 (2021).

[105] G. Vaisman, E. O. Kamenetskii, and R. Shavit, "Magnetic-dipolar-mode Fano resonances for microwave spectroscopy of high absorption matter", J. Phys. D: Appl. Phys. **48,** 115003(2015).

[106] G. Vaisman, E. Elman, E. Hollander, E. O. Kamenetskii, and R. Shavit, "Fano resonance microwave spectroscopy of high absorption matter", Patent: US 9651504 B2. 2017.